\documentclass[3p]{elsarticle}
\usepackage{lineno,hyperref}
\modulolinenumbers[5]

\journal{Int. J. Heat Mass Transf.}

\usepackage{graphicx}
\usepackage{epstopdf, epsfig}
\usepackage[section]{placeins} 
\usepackage{amsmath}
\usepackage{color}
\usepackage{rotating}

\newcommand{\remove}[1]{}

\biboptions{authoryear}

\begin{document}

\begin{frontmatter}

\title{Regimes of heat transfer in particle suspensions}


\author[Stoccolma]{Ali Yousefi\corref{mycorrespondingauthor}}
\cortext[mycorrespondingauthor]{Corresponding author: ayousefi@mech.kth.se}
\author[Stoccolma,Stanford]{Mehdi Niazi Ardekani}
\author[Padova]{Francesco Picano}
\author[Stoccolma,Norway]{Luca Brandt}

%

\address[Stoccolma]{Linn\' e FLOW Centre and SeRC (Swedish e-Science Research Centre), KTH, \\ Department of Engineering Mechanics, SE-10044 Stockholm, Sweden}
\address[Stanford]{Department of Chemical Engineering, Stanford University, Stanford, CA 94305, USA}
\address[Padova]{Department of Industrial Engineering, University of Padova, Via Venezia 1, 35131 Padova, Italy}
\address[Norway]{Department of Energy and Process Engineering, Norwegian University of Science and Technology (NTNU), Trondheim, Norway}
\begin{abstract}
We present results of interface-resolved simulations of heat transfer in suspensions of finite-size neutrally-buoyant spherical particles for solid volume fractions up to 35\% and bulk Reynolds numbers from 500 to 5600. An Immersed Boundary--Volume of Fluid method is used to solve the energy equation in the fluid and solid phase.
We relate the heat transfer to the regimes of particle motion previously identified, i.e.\ a viscous regime at low volume fractions and low Reynolds number, particle-laden turbulence at high Reynolds and moderate volume fraction and particulate regime at high volume fractions. We show that in the viscous dominated regime, the heat transfer is mainly due to thermal diffusion with enhancement due to the particle-induced fluctuations. In the turbulent-like regime, we observe the largest enhancement of the global heat transfer, dominated by the turbulent heat flux. In the particulate shear-thickening regime, however, the heat transfer enhancement decreases as mixing is quenched by the particle migration towards the channel core. As a result, a compact loosely-packed core region forms and the contribution of thermal diffusion to the total heat transfer becomes significant once again. The global heat transfer becomes, in these flows at volume fractions larger than 25\%, lower than in single phase turbulence.
\end{abstract}


\end{frontmatter}


\section{Introduction}\label{sec:introduction}

Heat transfer in wall-bounded particulate flows is commonly encountered in many industrial and environmental areas such as fuel combustion, food industry, pollution control and life science \citep{crowe2005multiphase,Guha2008}. In these flows, in addition to momentum/mechanical interactions between particles and fluid, the flow is also characterised by heat transfer between the two phases. The interactions among the two phases modulates the flow and result in modified wall drag and heat transfer rates \citep{ardekani-IJHFF-2018}. Therefore, predicting the heat transfer requires the knowledge of how particles are distributed across the domain, especially in relation to the wall, and of how particles affect the momentum transfer and finally how they affect the heat transfer within the suspension.  
Even though recent studies \citep{Deen2012,Tavassoli2013,Metzger2013,Sun2016,Esteghamatian2017,ardekani-JFM-2017,ardekani-IJHFF-2018,Lu2019moving} has shed some light on the matter, a comprehensive investigation of the heat transfer rate in finite-size particle suspensions at different Reynolds numbers and volume fractions is lacking in the literature. Here, we therefore perform particle-resolved direct numerical simulations (PR-DNS) of heat transfer in laminar and turbulent channel flows with neutrally buoyant, finite-size spherical particles up to $35\%$ volume fractions. We aim to quantify the heat-transfer process by breaking the total heat transfer into: (i) transport associated with the particle motion; (ii) convection by fluid velocity and (iii) thermal diffusion. 

\subsection{Particle suspension in wall-bounded flows}
As concerns wall-bounded particulate flows, \cite{Lashgari-PRL-2014,lashgari-IJMF-2016} documented the existence of three different regimes when changing the volume fraction $\phi$ of neutrally-buoyant spherical particles and the flow Reynolds number $Re$ (based on the flow bulk velocity): a laminar-like regime at low $Re$ and low to intermediate values of $\phi$, where the viscous stress dominates dissipation, a turbulent-like regime at high Reynolds number and low to intermediate $\phi$ where the turbulent Reynolds stress plays the main role in the momentum transfer across the channel and a third regime at higher $\phi$, denoted as inertial shear-thickening, characterised by a significant enhancement of the wall shear stress due to the particle-induced stresses. Indeed, thanks to novel and efficient numerical algorithms,  many studies have been dedicated in the recent years to the turbulence modulation in the presence of solid particles \citep{Lucci2010,Tanaka2015,Yu2016,Wang2016aa1,Gupta2018effect,Costa2020interface}. A decrease of the critical Reynolds number for transition to turbulence is reported by \cite{Matas2003,Loisel2013,Yu2013,Lashgari2015} for semi-dilute suspensions of neutrally-buoyant spherical particles,
consistent with an enhancement of the turbulence activity documented at low volume fraction (up to $10\%$) in turbulent flows \citep{picano-JFM-2015,Costa2016}. \cite{picano-JFM-2015} investigated dense suspensions in turbulent channel flow up to volume fraction of $20\%$. Their study revealed that the overall drag increase is due to the enhancement of the turbulence activity up to a certain volume fraction ($\phi \le 10\%$) and to significant particle-induced stresses at higher concentrations. \cite{Costa2016} explained that the turbulent drag increase in suspensions of spherical particles can be always partially
attributed  to the formation of a particle wall-layer, a layer of spheres forming near the wall in turbulent suspensions. Indeed, the particle wall-layer has been found to have a significant effect on the modulation of the near-wall turbulence, as confirmed in the case of  non-spherical particles \citep{Eshghinejadfard2017,Ardekani2019AR} where the absence of this layer leads to attenuation of the turbulence activity, resulting in drag reduction. \cite{picano-JFM-2015} attribute the formation of the near-wall layer of spherical particles to the strong wall-particle lubrication interaction that stabilizes the particle wall-normal position, forcing it to roll on the wall.  The particle dynamics in this layer alter the near-wall turbulence activity \cite{Ardekani2019turbulent} depending on the particle size and volume fraction. Following the work of \cite{Lashgari-PRL-2014}, we repeat the momentum budget analysis in this study, while considering smaller particles, and complete this with the analysis of the associated heat transfer. We will show that the onset of the particulate regime (inertial-shear thickening) shifts to higher volume fractions when reducing the particle size.

\subsection{Heat transfer}

Among recent studies in laminar and inertial shear-thickening regimes, \cite{Wang2009} presented experimental, theoretical and numerical investigations of the transport of fluid tracers between the walls bounding a sheared suspension of neutrally buoyant solid particles. These authors reported that the chaotic fluid velocity disturbances, caused by the suspended particles motion, lead to enhanced hydrodynamic diffusion across the suspension. In addition, it was found that for moderate values of the Peclet number,
 the Sherwood number, quantifying the ratio of the total rate of mass transfer to the rate of diffusive mass transport alone, changes linearly with the Peclet number. At higher Peclet numbers, however, the Sherwood number increases
 more slowly due to the increase in the mass transport resistance due to the presence of a molecular-diffusion boundary layer near the solid walls. 
 The effect of shear-induced particle diffusion on the transport of the heat across the suspension was investigated more recently by  \cite{Metzger2013} through a combination of experiments and simulations.  Their results further indicate that fluid velocity fluctuations due to the particle movement significantly increase the heat transfer through the suspension. 
  \cite{Souzy2015} investigated the mass transport in a cylindrical Couette cell of a sheared suspension with non-Brownian spherical particles and found that a rolling-coating mechanism (particle rotation convects the dye layer around the particles) transports convectively the dye directly from the wall towards the bulk.  The effect of particle inertia, volume fraction and thermal diffusivity ratio on the heat transfer in laminar Couette flow suspensions of spherical particles was investigated in \cite{ardekani-JFM-2017}.  This
  study revealed that inertia at the particle scale induces a non-linear increase of the heat transfer as a function of the volume fraction, whereas  it increases linearly at vanishing inertia \citep{Metzger2013}. The particle size and volume fraction effect on heat transfer in laminar pipe flows was studied in \citep{ardekani-IJHFF-2018}, where a considerable heat transfer enhancement was reported in the laminar regime by adding spherical particles. The heat transfer enhancement was shown to increase with the the pipe to particle diameter ratio and to saturate at high volume fractions ($40\%$).

  As concerns turbulent mutiphase flows, numerical algorithms coupling the heat and mass transfer are often challenging. Hence, in the first attempts, researchers used DNS only for the hydraulic characteristics of the flow and modelled the energy or mass transport equation \citep{Namburu2009,Chang2011,Bu2013,Liu2017}. Among earlier studies, \cite{Avila1995}, used a Lagrangian-stochastic-deterministic model (LSD) to show that high mass loadings of small particles increases the heat transfer rate, while at low mass loadings, the heat transfer rate decreases. Particle size effect is investigated in \cite{Hetsroni2002,Zonta2008}, who showed that larger particles increase the heat transfer coefficient more significantly than smaller ones by using a two-way coupling approximation. \cite{Kuerten2011} performed two-way coupling simulations of turbulent channel flow, showing an enhancement of the heat transfer and a small increase in the friction velocity in the presence of heavy small inertial particles with high specific heat capacity.  
\cite{ardekani-IJHFF-2018} studied the heat transfer within a suspension of neutrally buoyant, finite-size spherical particles in turbulent pipe flows, resolving also the temperature inside the rigid particles. These authors observed a transient increase in the heat transfer when increasing the particle volume fraction of the suspension, however, the process was reported to decelerate in time below the values in single-phase flows as high volume fractions of particles laminarize the core region of the pipe.

\subsection{Outline}
The governing equations, numerical method and the flow geometry are introduced in \S\ref{sec:Methodology}, followed by the results of the numerical simulations in section \S\ref{sec:results}. The main conclusions are finally drawn in \S\ref{sec:Final_remarks}. 
Here, we will consider 
finite-size particles at several volume fractions (up to $35\%$), from laminar to turbulent flows ($Re_b \in [500,5600]$) and relate the heat transfer properties to the flow regimes previously identified \citep{Lashgari-PRL-2014}. In particular, ensemble-average equations will be used to identify the different mechanisms contributing to heat transfer across the channel, namely molecular diffusivity and turbulent fluxes related to correlated fluid and particle motions.

\section{Methodology} \label{sec:Methodology}

\subsection{Governing equations}

We study a suspension of neutrally-buoyant rigid spherical particles, carried by a Newtonian fluid.
The incompressible Navier-Stokes equations describe the evolution of the carrier phase:
\begin{equation}
    \frac{\partial \boldsymbol{u}}{\partial t} + {\boldsymbol{u}\cdot \nabla \boldsymbol{u}} = - \frac{{\nabla} (p+p_e) }{\rho} + \nu \nabla^2 \boldsymbol{u} + \boldsymbol{f}\mathrm{,}
  \label{eq:Navier-Stokes}
\end{equation}
\begin{equation}
    {\nabla \cdot \boldsymbol{u}} = 0\mathrm{,}
  \label{eq:continuity}
\end{equation}
where $p$ is the modified fluid pressure (i.e. relative to the local hydrostatic load) and $\nabla p_e$ the imposed pressure gradient that drives the flow. $\nu$ is the kinematic viscosity of the fluid and $\rho$ the density of both the fluid and the particles.
$\boldsymbol{u} = (u,v,w)$ denotes the fluid velocity vector in the $(x,y,z)$ directions and the source term $\boldsymbol{f}$ accounts for the interactions between the carrier and the dispersed phase. Note that buoyancy effects due to density variations with the temperature are neglected in this work.

The motion of the rigid particles is governed by the Newton-Euler equations:

\begin{equation}
    m_p \frac{\mathrm{d} \boldsymbol{u}_p}{\mathrm{d} t} = \oint_{\partial \Omega_p} \boldsymbol{\boldsymbol{\tau} \cdot n} \,dA + \boldsymbol{F}_{c}\mathrm{,} 
    \label{eq:transtional}
\end{equation}
\begin{equation}
    I_p\frac{\mathrm{d}  \boldsymbol{\omega}_p}{\mathrm{d}t} = \oint_{\partial \Omega_p} \boldsymbol{r} \times (\boldsymbol{\boldsymbol{\tau} \cdot n}) \,dA + \boldsymbol{T}_{c}\mathrm{,}
    \label{eq:rotational}
\end{equation}
with $\boldsymbol{u_p}$ and $\boldsymbol{\omega_p}$ the particle linear and angular velocity vectors. 
$m_p$ and $I_p$ denote the particle mass and moment of inertia, $\boldsymbol{r}$ the position vector relative to the particle center and $\boldsymbol{n}$ the outward-pointing normal to the particle surface $\partial \Omega_p$.
$\boldsymbol{F}_c$ and $\boldsymbol{T}_c$ denote the force and torque resulting from the short-range particle-particle/wall interactions and the fluid stress tensor $\boldsymbol{\tau} = -p \boldsymbol{I} + \nu \rho ({\nabla} \boldsymbol{u} + {\nabla} \boldsymbol{u}^T)$.
The equations for the fluid and particle phase are coupled by the no-slip and no-penetration conditions on the particle surface, i.e. $\boldsymbol{u}|_{\partial \Omega} = \boldsymbol{u}_p + \boldsymbol{\omega}_p \times \boldsymbol{r}$.

The energy equation for an incompressible flow reads as:
\begin{equation}
    \frac{\partial T}{\partial t} + \boldsymbol{u \cdot \nabla} T = \boldsymbol{\nabla \cdot} (\alpha \boldsymbol{\nabla} T) \mathrm{,}
    \label{eq:energy}
\end{equation}
where $T$ is the temperature and $\alpha$ the thermal diffusivity, equal to $k/ (\rho C_p)$, with $k$ the thermal conductivity and $C_p$ the specific heat capacity. 
In the present work, the same thermal diffusivity is considered for the fluid and particles and Eq.~\eqref{eq:energy} is resolved on every grid point in the computational domain so to fully resolve the heat transfer also inside the particles.

\subsection{Numerical method}

We use the direct-forcing immersed boundary method (IBM), initially developed by \cite{uhlmann-JCP-2005} and modified by \cite{breugem-JCP-2012}, to fully resolve fluid-solid interactions.
A volume of fluid (VoF) approach \citep{hirt-JCP-1981} is coupled with the IBM to solve Eq.~\eqref{eq:energy} in the two phases \citep{strom-IJMF-2013}.
The method has been used extensively with several validations reported by \cite{picano-JFM-2015}, \cite{lashgari-IJMF-2016} and \cite{ardekani-IJMF-2016}, considering fluid-solid interactions and by \cite{ardekani-JFM-2017}, \cite{ardekani-IJHFF-2018} and \cite{majlesara-IJHMT-2020} for the heat transfer in particle-laden flows. All the details of this implementation are presented in the mentioned references; for the sake of completeness, only a brief description of the method is therefore presented here.

The Navier-Stokes equations governing the fluid phase are solved on a uniform $(\Delta x = \Delta y = \Delta z)$, staggered, Cartesian grid. The spherical particles are discretized by a set of Lagrangian points, uniformly distributed along their surface.
The IBM forcing scheme consists of three steps: (i) the fluid prediction velocity is interpolated from the Eulerian to the Lagrangian grid, (ii) the IBM force required for matching the local fluid velocity and the local particle velocity is computed on each Lagrangian grid point and (iii) the resulting IBM force is spread from the Lagrangian to the Eulerian grid. The interpolation and spreading operations are done through the regularized Dirac delta function of \cite{roma-JCP-1999}, which acts over three grid points in all coordinate directions.

When the gap between two particles (or a particle and the wall) is smaller than the grid spacing, the IBM fails to resolve the short-range hydrodynamic interactions. Therefore, we use a lubrication correction model based on the asymptotic analytical expression for the normal lubrication force between two equal spheres \citep{brenner-CES-1961}.
When the particles are in collision, the lubrication force is turned off and a collision force based on the soft-sphere model is activated. The restitution coefficients, used for normal and tangential collisions, are $0.97$  and $0.1$, with Coulomb friction coefficient $0.15$.
More details on the short-range models and corresponding validations can be found in \cite{costa-PRE-2015}. 

Using the volume of fluid (VoF) approach, proposed in \cite{ardekani-JFM-2017}, the velocity of the combined phase is defined at each point in the domain as
\begin{eqnarray}
\label{eq:Ucp}  
\textbf{u}_{cp} &=&  \left( 1 - \xi \right) \textbf{u} \, + \, \xi \textbf{u}_p , \, 
\label{eq:ALPHAcp}
\end{eqnarray} 
where $\textbf{u}$ is the fluid velocity and $\textbf{u}_p$ the solid phase velocity, obtained by the rigid body motion of the particle at the desired point. In other words, the fictitious velocity of the fluid phase trapped inside the particles is replaced by the particle rigid body motion velocity when solving the temperature equation inside the solid phase; this velocity is  computed as $\textbf{u}_p + \pmb{\omega}_p \times \textbf{r}$ with $\textbf{r}$, the position vector from the center of the particle. $\xi$ is a phase indicator, obtained from the location of the fluid/solid interface exactly for rigid spheres and used to distinguish the solid and the fluid phase within the computational domain. $\xi$ is computed at the velocity (cell faces) and the pressure points (cell center) throughout the staggered Eulerian grid. This value varies between $0$ and $1$ based on the solid volume fraction of a cell of size ${\Delta x}$ around the desired point. $\textbf{u}_{cp}$ is then used in Eq.~\eqref{eq:energy} where the same thermal diffusivity in both phases is considered. It should be noted that the computed $\textbf{u}_{cp}$ remains a divergence free velocity field. 

Eq.~\eqref{eq:energy} is discretized around the Eulerian cell centers (pressure and temperature points on the Eulerian staggered grid) and integrated in time, using the same explicit low-storage Runge-Kutta method \citep{Wesseling2009}, employed for the flow solver.
Spatial derivatives are estimated with the second order central-difference scheme, except for the advection term $\textbf{u}_{cp} \cdot \nabla T$, where the explicit fifth-order weighted essentially non-oscillatory (WENO) scheme \citep{Liu1994,Sugiyama2011,Rosti20171} is used to avoid dispersive behaviors of the temperature field.  

\subsection{Computational setup}

In the present work, we investigate the channel flow between two infinite parallel walls, laden with rigid spherical particles. 
The size of the computational domain is $L_x = 6h$, $L_y = 2h$ and  $L_z = 3h$ in the streamwise, wall-normal and spanwise directions, where $h$ is half the distance between the channel walls.
The flow is periodic in the streamwise and spanwise directions, with no-slip and no-penetration boundary conditions imposed at the bottom and top boundaries.
A uniform pressure gradient $(\nabla p_e)$ forces the flow such that the bulk fluid velocity $U_b$ remains constant.
The non-dimensional temperature is defined as $T^{\ast} \equiv (T - T_{cold}) / (T_{hot} - T_{cold})$, where $T_{hot}$ and $T_{cold}$ are the operating temperatures at the walls. Therefore, the non-dimensional temperatures $T^{\ast}$ for the upper and lower walls are fixed at $T^{\ast} = 0$ and $1$.

In the simulations we examine the heat transfer efficiency when varying the bulk Reynolds number defined as $Re_b = (2h U_b) / \nu$, the thermal diffusivity, i.e. the Prandtl number defined as $Pr = \nu / \alpha$ and the solid volume fraction $\phi$, defined as the total volume of solid particles over the total volume of the computational domain.
Details of the parameters used in the simulations are given in table~\ref{tab:parameters}.
The bulk Reynolds numbers considered here span over laminar and turbulent regimes. To fully resolve the flow, for the cases with $Re_b \leq 4000$, am Eulerian grid resolution of $24/D$ grid points is used, with $D$ the particle diameter; for the cases with $Re_b=5600$, instead, we use a higher resolution, $32/D$ grid points.
This corresponds to maximum grid spacing of $\Delta x ^+ = u_{\tau} \Delta x / \nu = 0.74$ in wall units of the single-phase flow case with $Re_b=5600$ and guarantees that all the flow scales are resolved; here, $u_{\tau} = \sqrt{\tau_{tot} / \rho}$ is the friction velocity with $\tau_{tot}$ the total stress, i.e. the sum of the viscous and the Reynolds stress.
The diameter of the particles considered in this study is such that $2h/D = 15$, which corresponds approximately to $D^+ = u_{\tau} D / \nu = 24$ in wall units of the single-phase flow case with $Re_b=5600$.
The number of Lagrangian grid points $N_L$ on the surface of each particle is defined such that the Lagrangian grid volume $\Delta V_l$ becomes equal to the volume of the Eulerian mesh $\Delta x^3$.
This translates to a total number of $N_L = 1721$ and $3219$ Lagrangian points for the cases with Eulerian grid resolution of $24/D$ and $32/D$, uniformly distributed on the surface of each particle.

The simulations are started with a random distribution of particles inside the domain and a linear temperature field $T^{\ast}$ between $0$ and $1$ for both the fluid and the particles.
Statistics are collected over an interval of almost $\Delta t U_b /D = 2000$  after the wall-normal heat flux has reached a steady-state value with small oscillations in time.

\begin{table}[t]
    \begin{center}
    \def~{\hphantom{0}}
    \begin{tabular}{lcc}
        \hline
        $Re_b$        & $\phi$                                & $Pr$     \\[3pt]
        \hline
        $500$         & $1,5,10,20,30   \,  \& \,  35\%$      & $1,4,7$  \\
        $1000$        & $1,5,10,20,30   \,  \& \,  35\%$      & $1,4,7$  \\
        $2000$        & $1,5,10,20,30   \,  \& \,  35\%$      & $1,4,7$  \\
        $3000$        & $0,1,5,10,20,30 \,  \& \,  35\%$      & $1,4,7$  \\
        $4000$        & $0,1,5,10,20,30 \,  \& \,  35\%$      & $1,4,7$  \\
        $5600$        & $0,1,5,10,20,30 \,  \& \,  35\%$      & $1,4,7$  \\
        \hline
    \end{tabular}
    \caption{Parameters of the DNS data set; $Re_b$ denotes the bulk Reynolds number, $\phi$ the solid volume fraction and $Pr$ the Prandtl number.}
    \label{tab:parameters}
    \end{center}
\end{table}

\section{Results}\label{sec:results}

\begin{figure}
  \centering
   \includegraphics[width=0.95\textwidth]{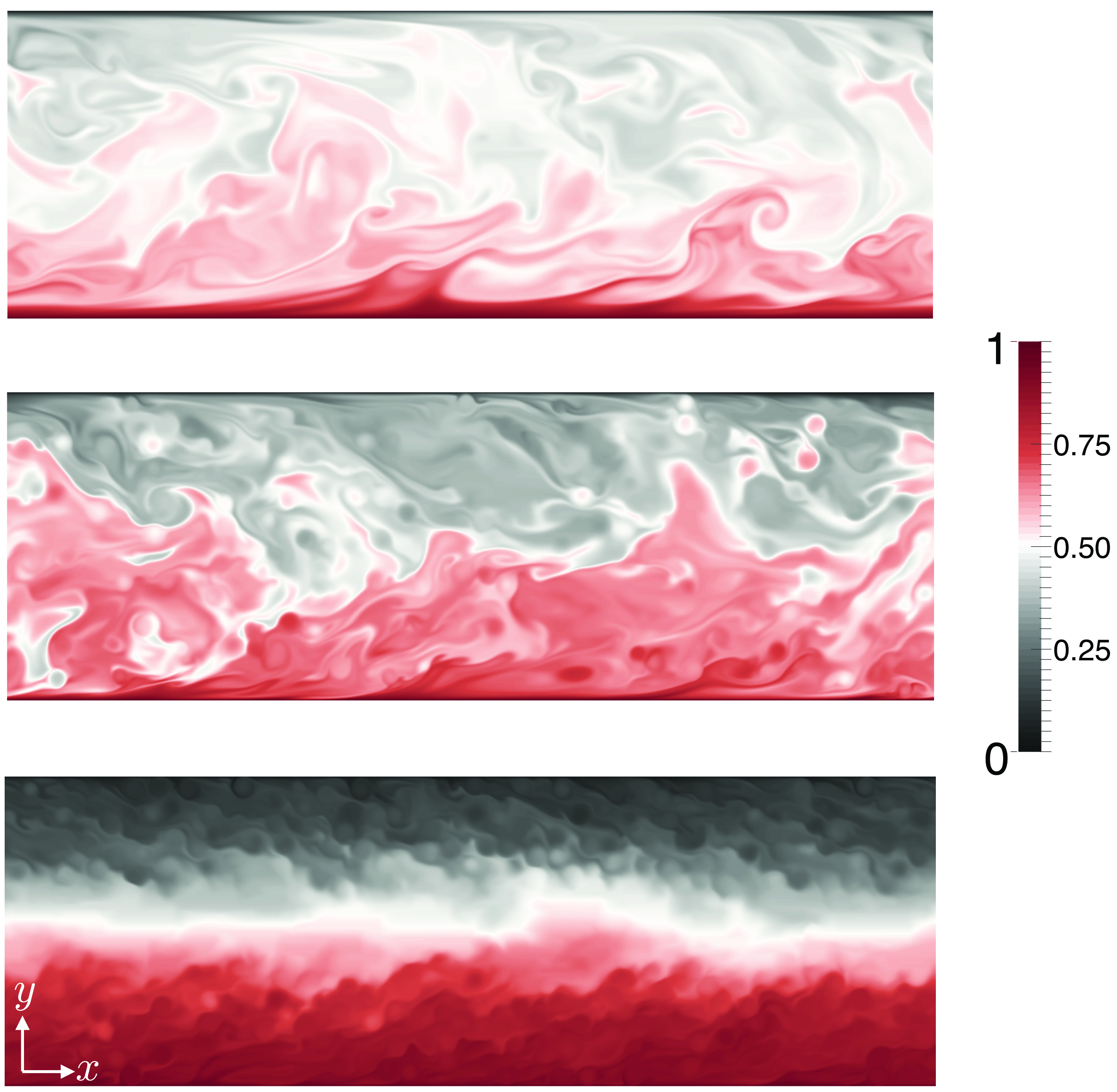}
   \put(-40,315){$T^{\ast}$}
   \put(-460,427){$(a)$}
   \put(-460,275){$(b)$}
   \put(-460,120){$(c)$}\\
   \caption{Instantaneous contours of the temperature in a wall-normal $x\mbox{-}y$ plane at the non-dimensional time $tU_b/D = 514$ for $Re_b=5600$ and $Pr=7$ for different solid volume fractions: (a) $\phi=0\%$, (b) $\phi=10\%$ and (c) $\phi=35\%$.}
\label{fig:snapshots}
\end{figure}
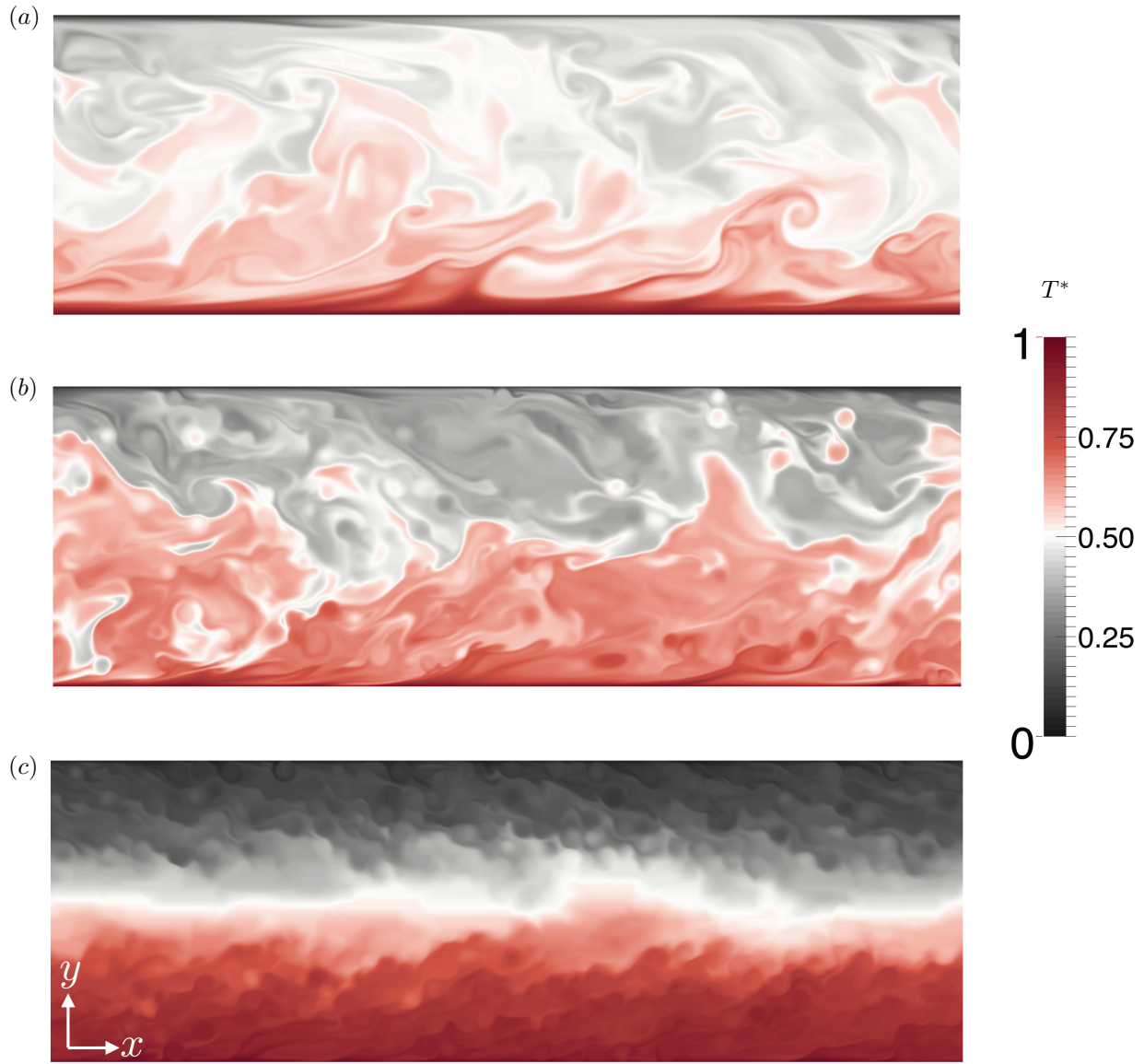

We start by showing snapshots of the temperature field in the suspension flow, see figure~\ref{fig:snapshots}, where the instantaneous contours of the temperature in a wall-normal $x\mbox{-}y$ plane are depicted for the volume fractions: (a) $\phi=0\%$, (b) $\phi=10\%$ and (c) $\phi=35\%$ at $Re_b=5600$ and $Pr=7$.
The contours are plotted at the non-dimensional time $tU_b/D = 514$, at which all the cases have passed the initial transient state and reached the fully-developed turbulent and heat transfer state.
The figure shows that the mixing in the temperature field observed for the single-phase case, panel (a), is more intense for the volume fraction $\phi=10\%$, panel (b), where traces of \textit{hot} particles in the \textit{cold} fluid regions and visa versa are clearly observable.
Conversely, for the volume fraction $\phi=35\%$, cf.\ panel (c), the layering of the particles is evident with low wall-normal motions across the channel.

\begin{figure}
  \centering
   \includegraphics[width=0.495\textwidth]{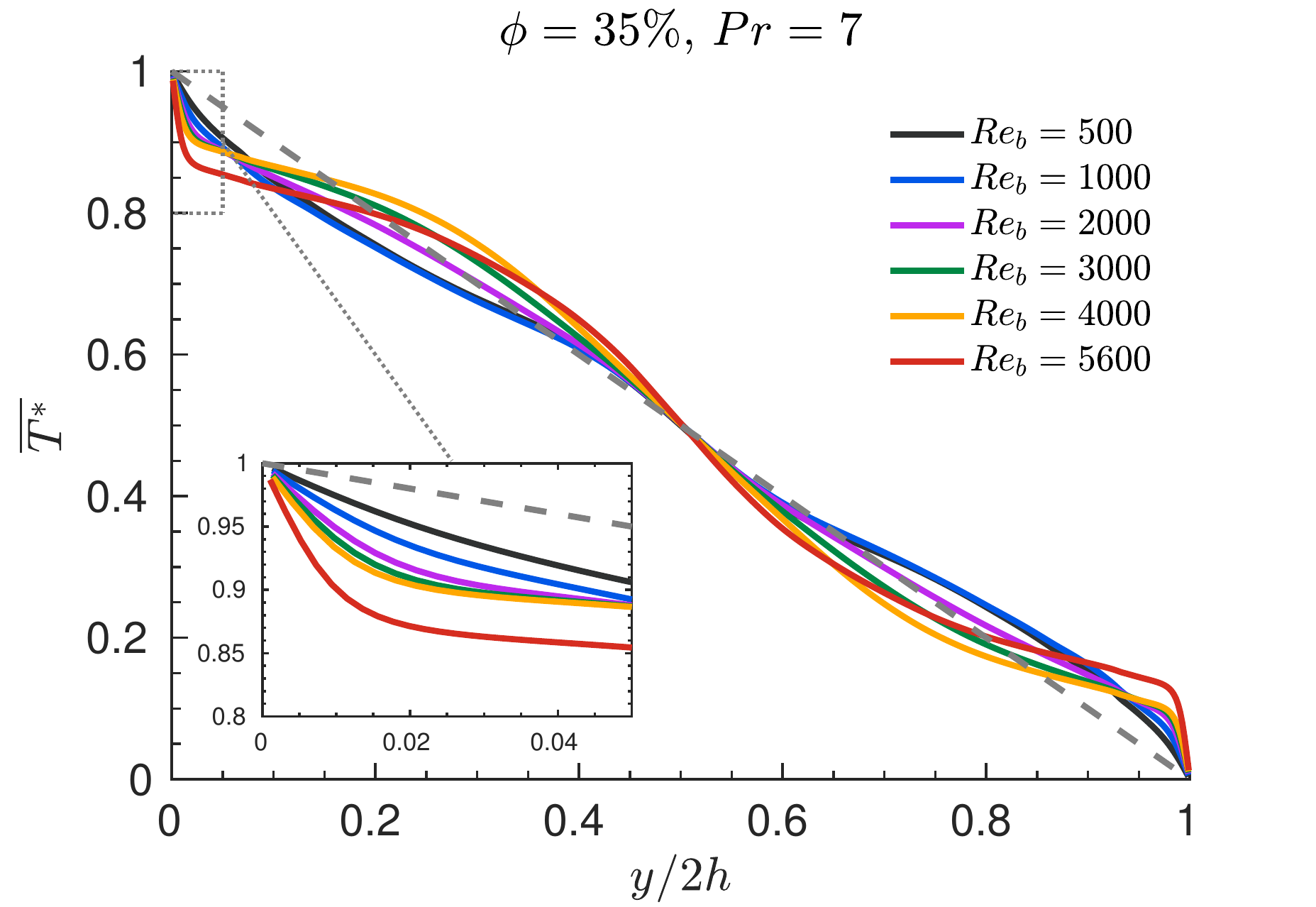}
   \includegraphics[width=0.495\textwidth]{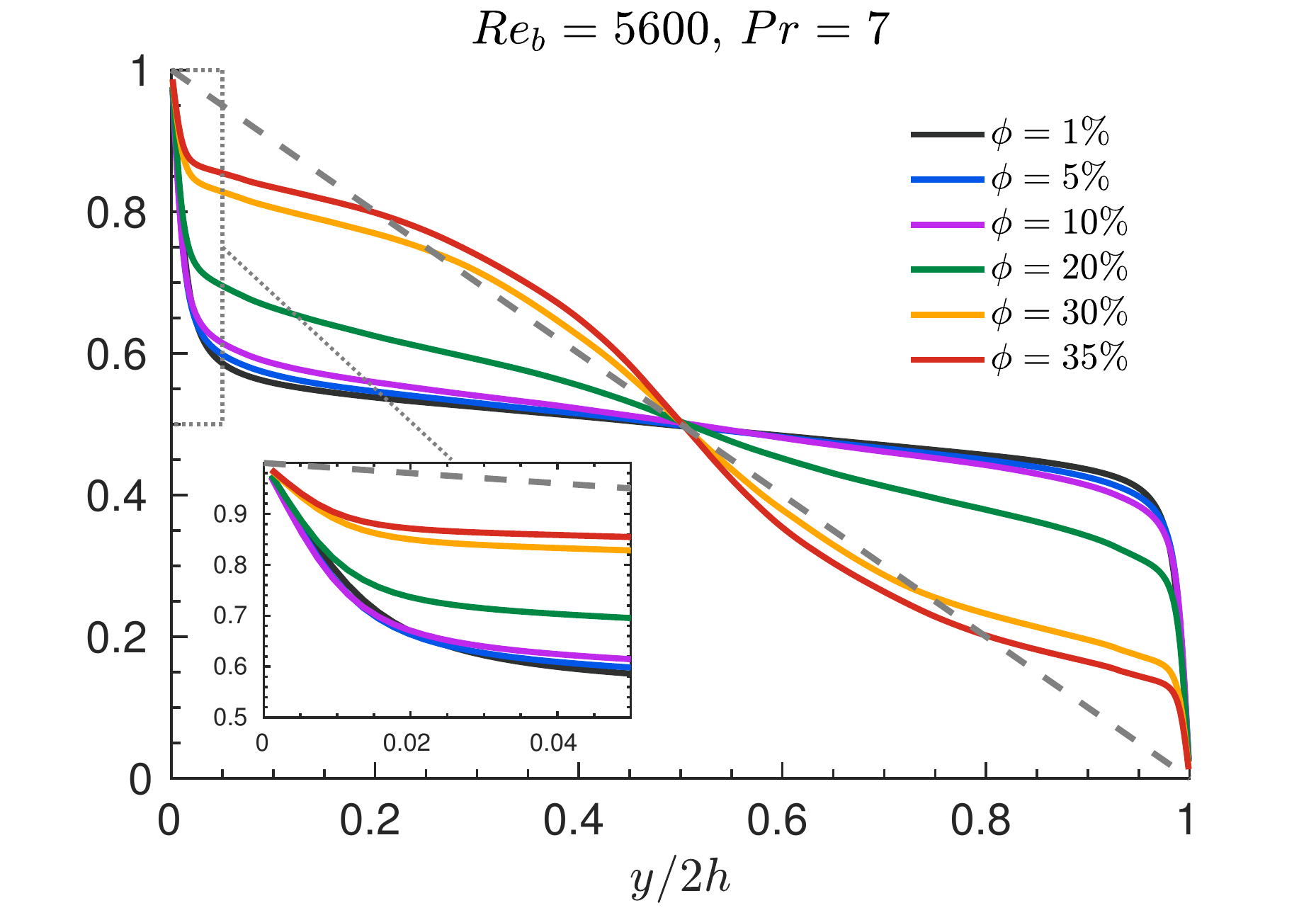}
   \put(-470,150){$(a)$}
   \put(-230,150){$(b)$}\\
   \caption{Mean temperature, $\overline{T^{\ast}}$, wall-normal profiles at $Pr=7$ for: (a) $\phi=35\%$ and different Reynolds numbers and (b) $Re_b=5600$ and different values of the solid volume fraction $\phi$. The insets show the magnified view close to the bottom wall. The dashed gray lines denote the linear laminar single-phase profile.}
\label{fig:mean_temperature}
\end{figure}

Figure\ref{fig:mean_temperature} shows the mean temperature, $\overline{T^{\ast}}$, wall-normal profiles for a number of cases: panel(a) displays the results for different Reynolds numbers and $\phi=35\%$, while panel (b) data for different volume fractions at $Re_b=5600$. 
By increasing $Re_b$ (panel a), the temperature gradient at the walls, i.e.\ the total heat transfer, increases. Away from the walls, the temperature gradient is closer to that of the linear laminar profile.
On the other hand, at $Re_b=5600$, the temperature gradient at the walls increases first and then decreases when increasing the solid volume fraction. Away from the walls, the cases with $\phi \leq 20\%$ display small values of the temperature gradient, indicating good mixing, while approaching the linear  profile when $\phi \geq 30\%$.
In the following, we will quantify the effect of the particle concentration and inertia, i.e.\ the Reynolds number, on the overall heat transfer efficiency and the mechanisms of heat transfer across the channel.

\subsection{Heat transfer enhancement}

To quantify the effect of each parameter on the heat transfer between the two walls, we will use the Nusselt number, $Nu$, defined as:  
\begin{equation}
    Nu \equiv \frac{\mathcal{H} 2h}{k} = \frac{d T^{\ast}}{dy^{\ast}}\bigg|_{y^{\ast}=0} \mathrm{,}
    \label{eq:Nusselt}
\end{equation}
where $\mathcal{H}$ is the average heat transfer coefficient and $y^{\ast} \equiv y/2h$ the distance from the wall, normalized by the channel height.

\begin{figure}
  \centering
   \includegraphics[width=0.495\textwidth]{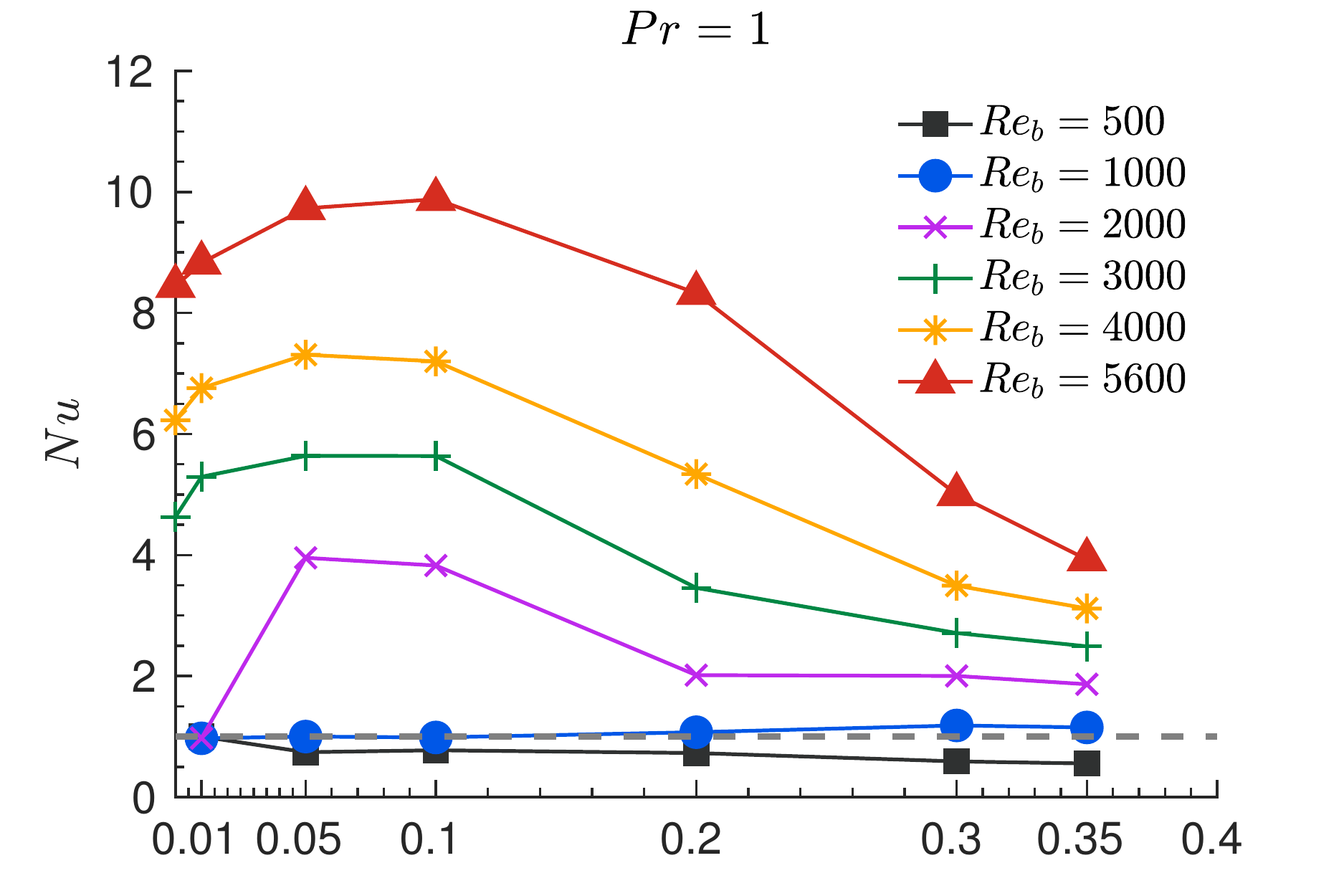}
   \includegraphics[width=0.495\textwidth]{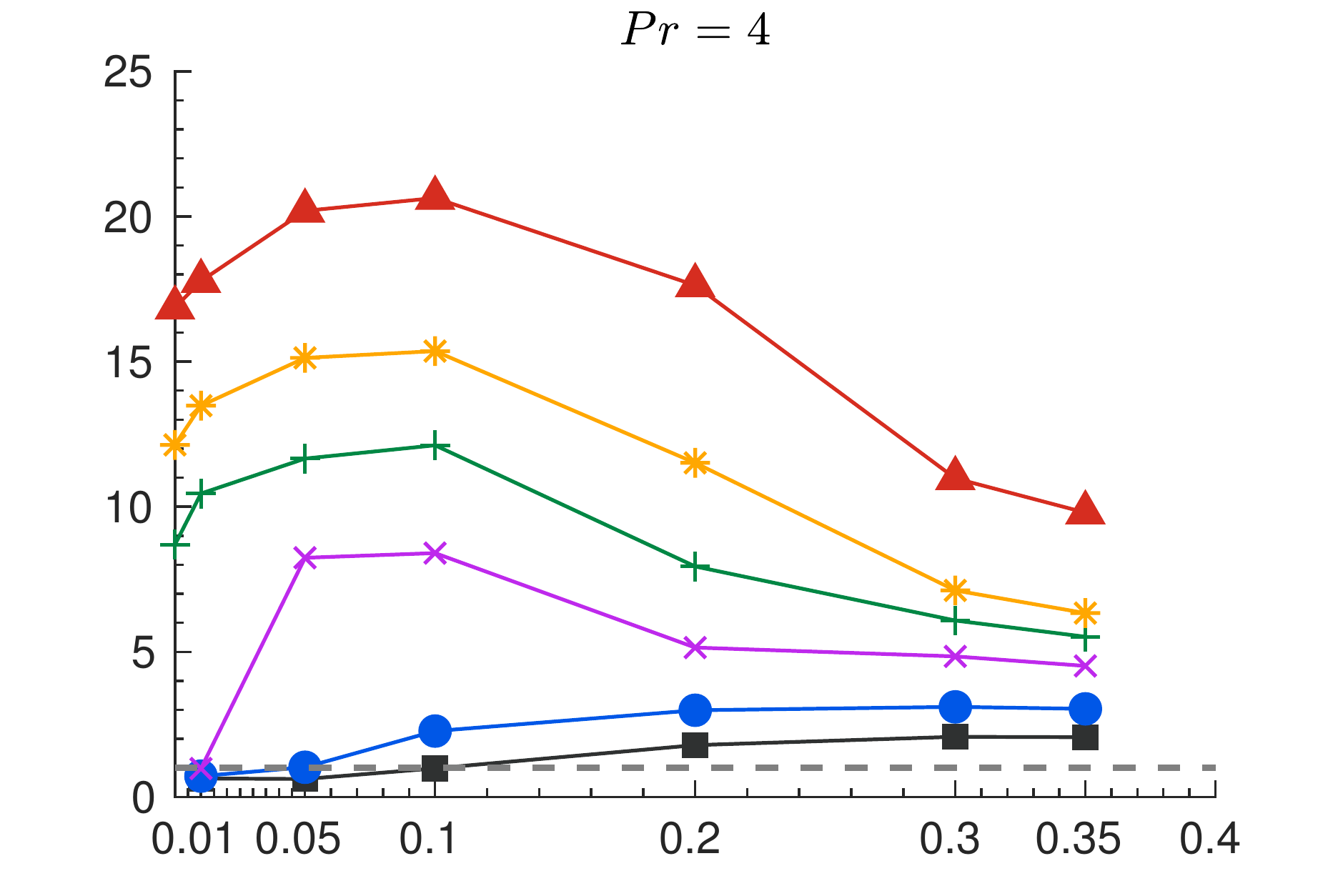}
   \put(-470,150){$(a)$}
   \put(-230,150){$(b)$}\\
   \includegraphics[width=0.495\textwidth]{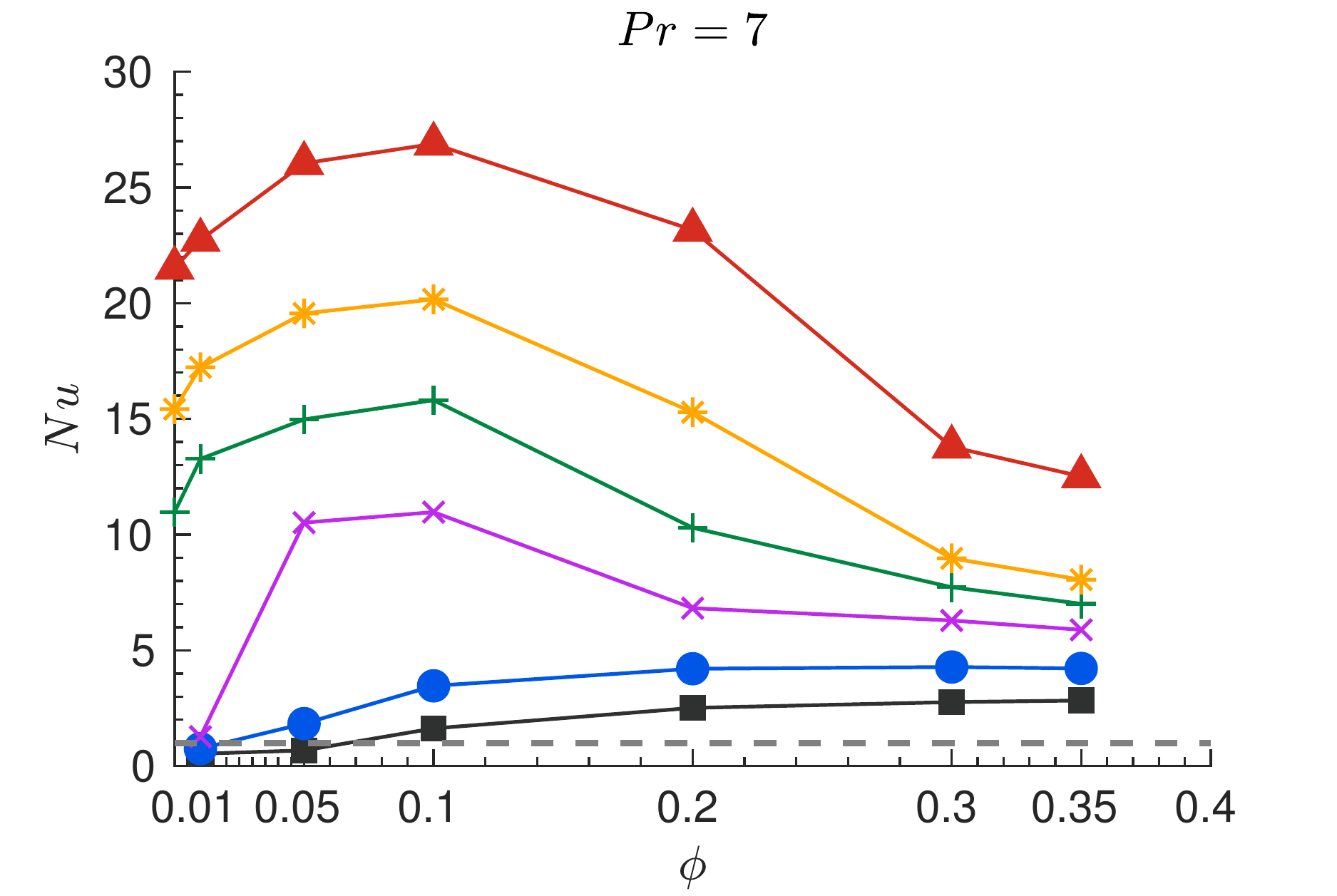}
   \includegraphics[width=0.495\textwidth]{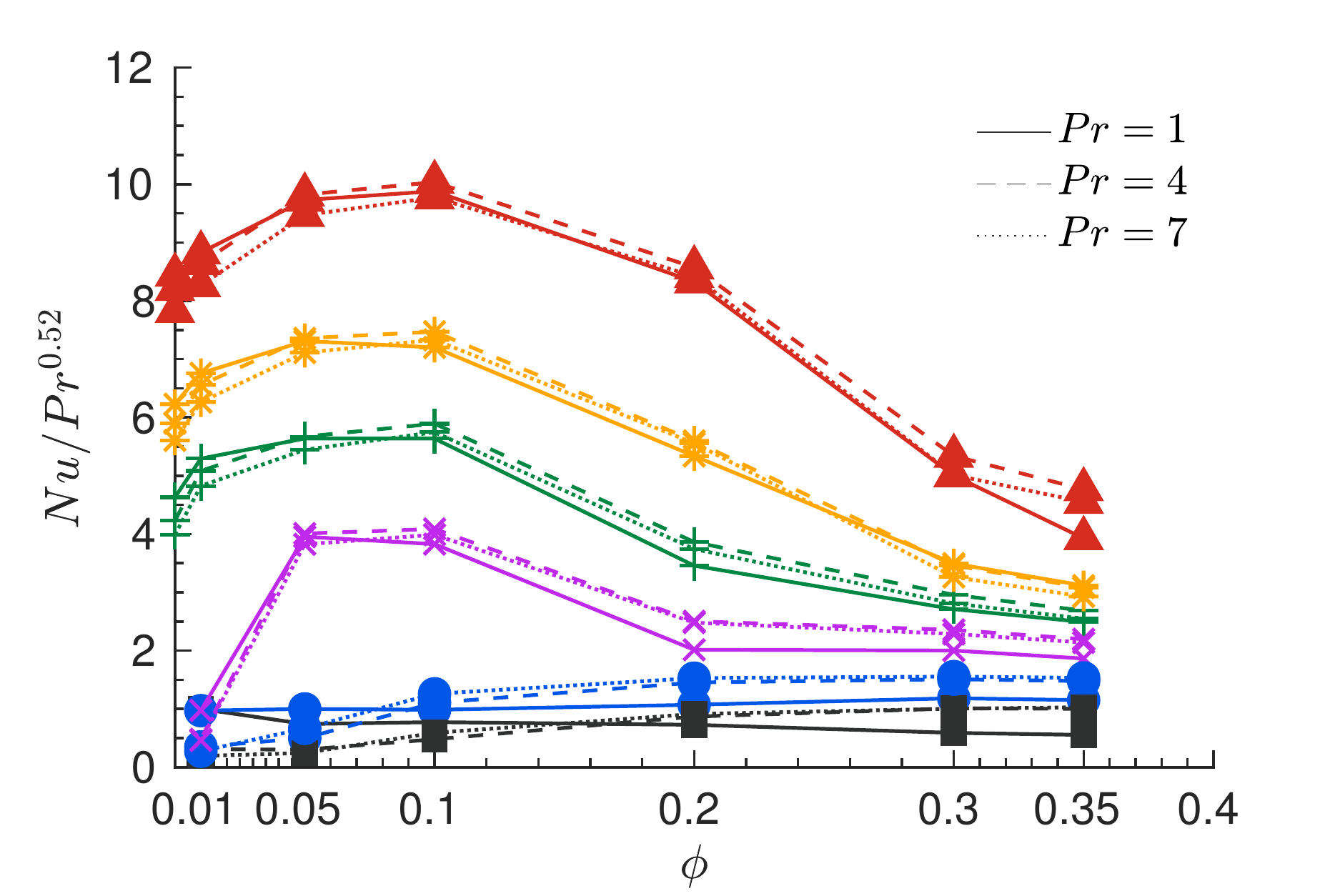}
   \put(-470,150){$(c)$}
   \put(-230,150){$(d)$}\\
   \includegraphics[width=0.8\textwidth]{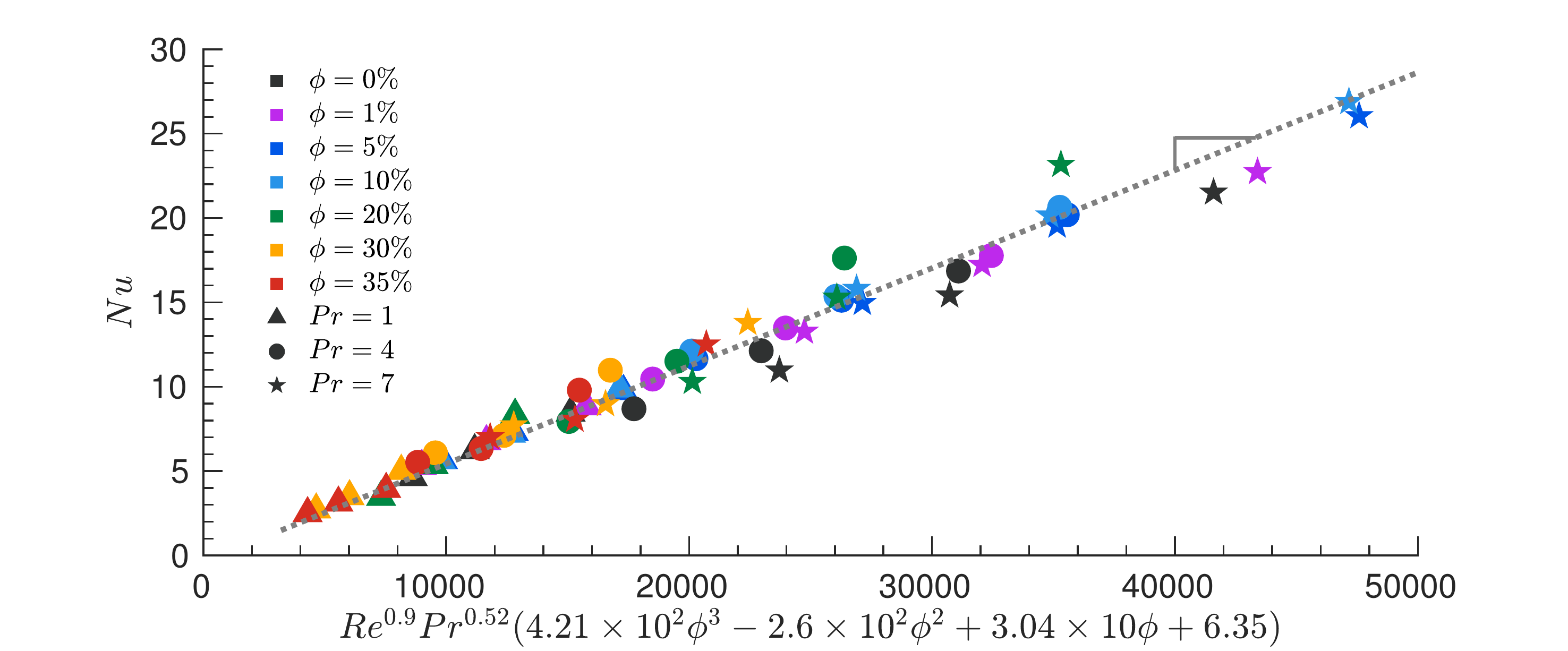}
   \put(-99,117){\scriptsize  $1$}
   \put(-92,125){\scriptsize  $1725$}
   \put(-355,150){$(e)$}\\
   \caption{The Nusselt number, $Nu$, versus the volume fraction $\phi$ for different Reynolds numbers; (a)$Pr=1$, (b)$Pr=4$ and (c)$Pr=7$. The dashed gray lines in panels (a-c) denote the value $Nu=1$. Panel (d) shows the value $Nu/Pr^{0.52}$ as a function of the solid volume fraction $\phi$ for all the cases. Panel (e) shows the Nusselt number obtained using Eq.~\ref{eq:correlation} for the cases with $3000 \leq Re_b \leq 5600$; the dotted gray line denotes the least-square regression line.}
\label{fig:heat_enhancement}
\end{figure}

The Nusselt number, $Nu$, is reported in figure~\ref{fig:heat_enhancement} as a function of the volume fraction $\phi$ for different Reynolds numbers and for the different values of the Prandtl number under investigations: panel (a) $Pr=1$, (b) $Pr=4$ and (c) $Pr=7$.
For the cases with $Re_b=500$ and $1000$, in the laminar flow regime, we observe a monotonic increase of the Nusselt number, increasing the volume fraction, with a gradual saturation for concentrations higher than $\phi=30\%$; the exception is the cases with $Re_b=500$ and $Pr=1$, where $Nu$ slightly decreases, increasing the volume fraction.
The increase with inertia is in-line with the results of \cite{ardekani-IJHFF-2018} in pipe flows with similar relative particle size, i.e.\ pipe diameter to particle diameter, and $Re_b=370$.
These authors showed that the heat transfer enhancement reduces significantly as the ratio between the pipe to the particle diameter increases, which is associated to the reduction of inertial effects at the particle scale.
Here, we increase inertial effects by increasing  $Re_b$ from $500$ to $1000$, which also results in enhanced heat transfer, i.e.\ larger values of $Nu$. 

At $Re_b=2000$, in the transitional regime between laminar and turbulent flow, the heat transfer efficiency increases significantly when increasing the volume fraction from $1\%$ to $10\%$, then decreases until $\phi=20\%$ and saturates beyond that.
It should be noted that at this bulk Reynolds number, the noise introduced by the presence of the particles can trigger the transition from laminar to the turbulent regime.
In this case, the flow is turbulent-like at volume fractions of the order of 10\%, whereas it behaves more similarly to a laminar flow again at higher $\phi$ as the flow cannot sustain strong fluctuations in the presence of  larger dissipation induced by many particles, see figure~\ref{fig:RSS}(a).

For $Re_b \geq 3000$, we observe a clear increase of the Nusselt number, which is an indication of the transition  to the turbulent regime even in the absence of particles. 
Increasing the solid volume fraction $\phi$, the Nusselt number, $Nu$ increases till $\phi=10\%$ and then decreases, reaching values below those of the turbulent single-phase cases at the same $Re_b$ for $\phi > 20\%$.

The Prandtl number $Pr$ quantifies the ratio of the momentum diffusivity to thermal diffusivity; hence, increasing $Pr$ from $1$ to $7$ in our study, the relative speed of the heat transfer via conduction, which is the only mechanism of heat transfer between the walls in the single-phase laminar flow, decreases and the diffusion of momentum is comparatively faster.
As a consequence, 
the enhancement of the heat transfer coefficient, or the Nusselt number, increases with the Prandtl number for each value of the solid volume fraction $\phi$ and  bulk Reynolds number $Re_b$, as shown in figure~\ref{fig:heat_enhancement} (a-c): 
in other words, the heat transfer enhancement is more important when increasing the Prandtl number as this is related to the diffusion of fluid momentum induced by the particle motion and by the turbulence.

To solely look at the effect of the particle concentration and the bulk Reynolds number on the heat transfer efficiency, we normalize the Nusselt number as $Nu/Pr^{0.52}$ for each case in figure~\ref{fig:heat_enhancement} (d). The figure shows that by this scaling the data collapse fairly well for all the cases, especially in the turbulent regime. 
In the literature, proposed semi-empirical correlations between the Nusselt number and the Prandtl number for various geometries in turbulent single-phase flows have the form of $Nu \propto Pr^a$ with $a \in [0.3-0.6]$, see e.g. \cite{yeh1984review,vajjha2015development,taler2016new}.
Interestingly, the mentioned scaling works well independent of the volume fraction also for our results. This indicates that the underlining mechanisms work in a similar way, though the origin is different (particle versus turbulent transport).

For the turbuelent cases, i.e. $3000 \leq Re_b \leq 5600$, we derive a correlation between the Nusselt number, the bulk Reynolds number and the volume fraction, using polynomial curve-fitting.
    Figure~\ref{fig:heat_enhancement} (e) shows the Nusselt number, using the obtained correlation function, versus the simulation results.
Fitting a least-square regression line, the overall correlation function  $Nu=Nu(Re_b, Pr, \phi)$ reads as:
\begin{multline}
    Nu = 5.79 \times 10^{-4} \times [Re_b^{0.9} Pr^{0.52} (4.21 \times 10^2 \phi^3 - 2.61 \times 10^2 \phi^2 + 3.04 \times 10 \phi + 6.35)] \, \mathrm{;} \\
      3000 \leq  Re_b \leq 5600 \mathrm{,} \, \, \, 1 \leq Pr \leq 7 \mathrm{,} \, \, \, 0\% \leq \phi \leq 35\% \mathrm{.} 
\label{eq:correlation}
\end{multline}

\subsection{Heat transfer budget}

To gain further insight into the heat transfer modifications presented above, we examine the contributions of the different mechanisms responsible for the total heat flux between the channel walls. Following the phase-ensemble average presented in \cite{ardekani-JFM-2017}, we write the wall-normal heat flux as
\begin{equation}
    q_{tot}^{\prime \prime} = q_{C_p}^{\prime \prime} + q_{C_f}^{\prime \prime} + q_{D}^{\prime \prime} \mathrm{,}
    \label{eq:heat_budget1}
\end{equation}
\begin{equation}
    q_{C_p}^{\prime \prime} = - \Phi \langle v_p^{\prime} T_p^{\prime} \rangle \mathrm{,}
    \label{eq:heat_budget2}
\end{equation}
\begin{equation}
    q_{C_f}^{\prime \prime} = - (1 - \Phi) \langle v_f^{\prime} T_f^{\prime} \rangle \mathrm{,}
    \label{eq:heat_budget3}
\end{equation}
\begin{equation}
    q_{D}^{\prime \prime} = \alpha  \frac{dT}{dy} \mathrm{,}
    \label{eq:heat_budget4}
\end{equation}
where $\Phi$ is the local volume fraction, $v^{\prime}$ and $T^{\prime}$ the wall-normal velocity and temperature fluctuations with subscripts $_f$ and $_p$ denoting the fluid and particle phases respectively; $\langle \cdot \rangle$ indicates the ensemble average conditioned to the considered phase.
The total heat flux $q_{tot}^{\prime \prime}$ is divided into three terms: (i) convection by the correlation of the particle velocity and temperature fluctuations  $q_{C_p}^{\prime \prime}$, (ii) convection by the correlation of the fluid velocity and temperature fluctuations $q_{C_f}^{\prime \prime}$ and (iii) molecular diffusion $q_D^{\prime \prime}$; since we consider the same thermal diffusivity inside the fluid and particles, we show the contributions by the molecular diffusion in the solid and fluid phases together as one term.

\begin{figure}
  \centering
   \includegraphics[width=0.495\textwidth]{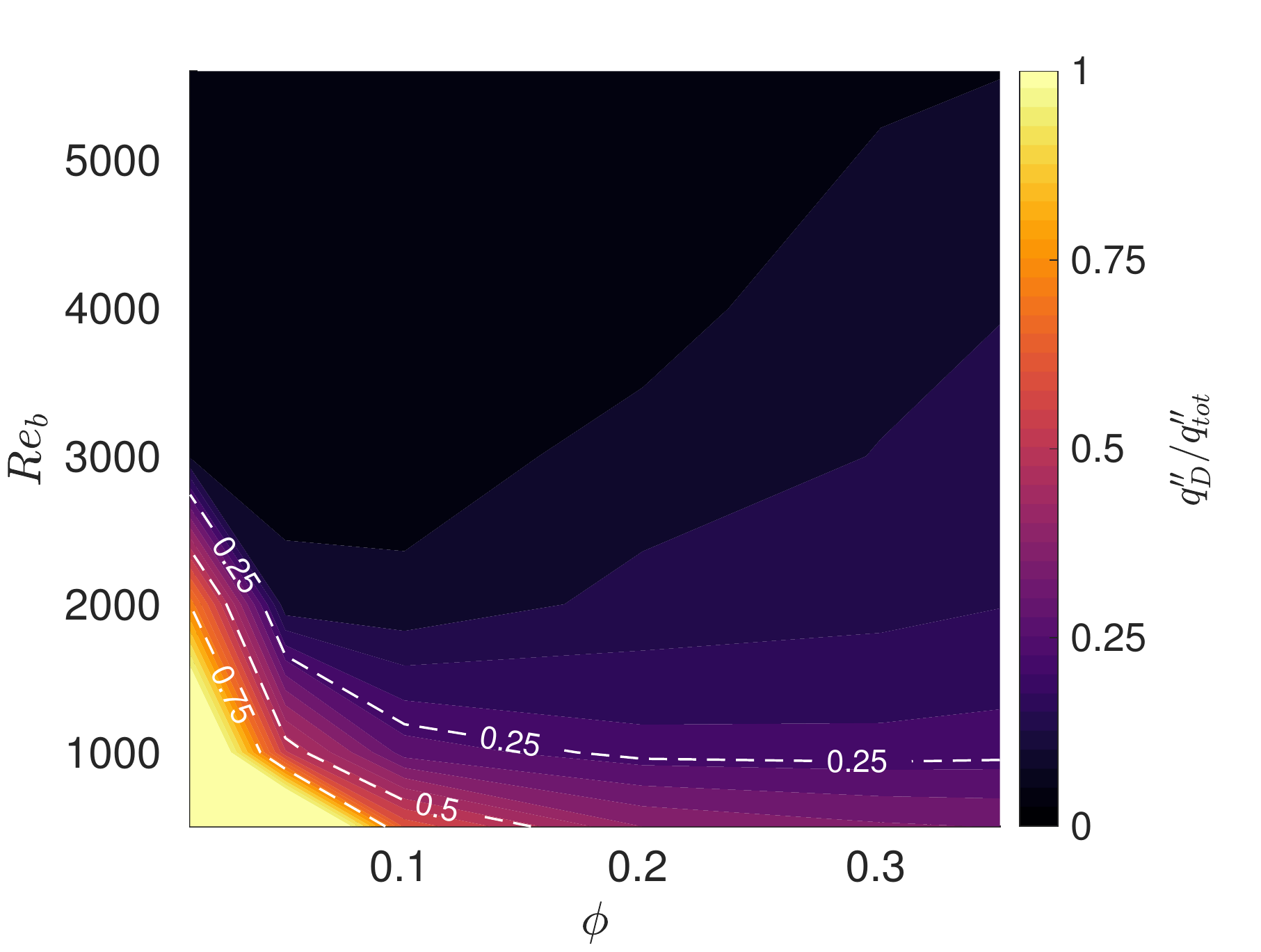}
   \includegraphics[width=0.495\textwidth]{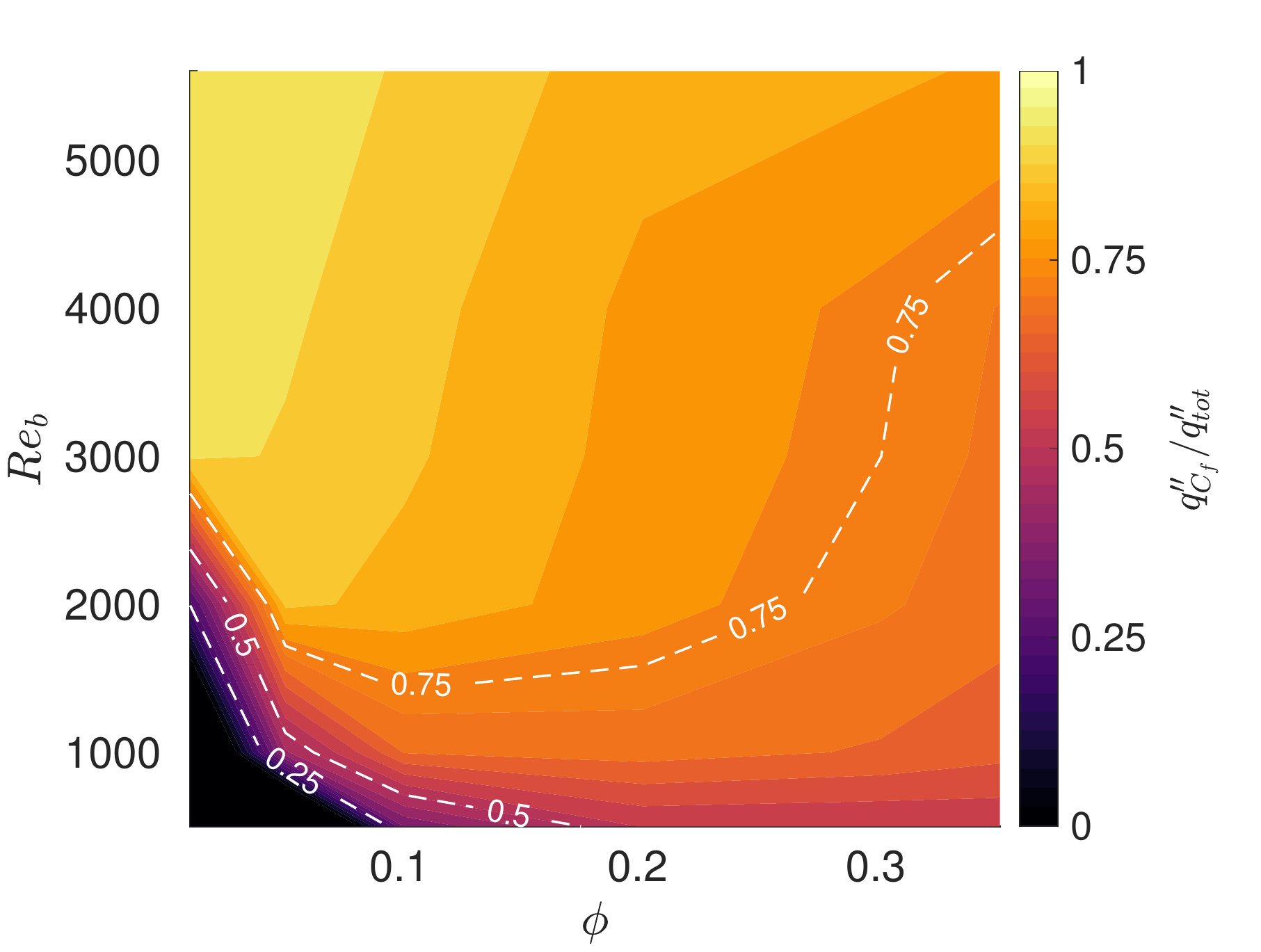}
   \put(-470,160){$(a)$}
   \put(-230,160){$(b)$}\\
   \includegraphics[width=0.495\textwidth]{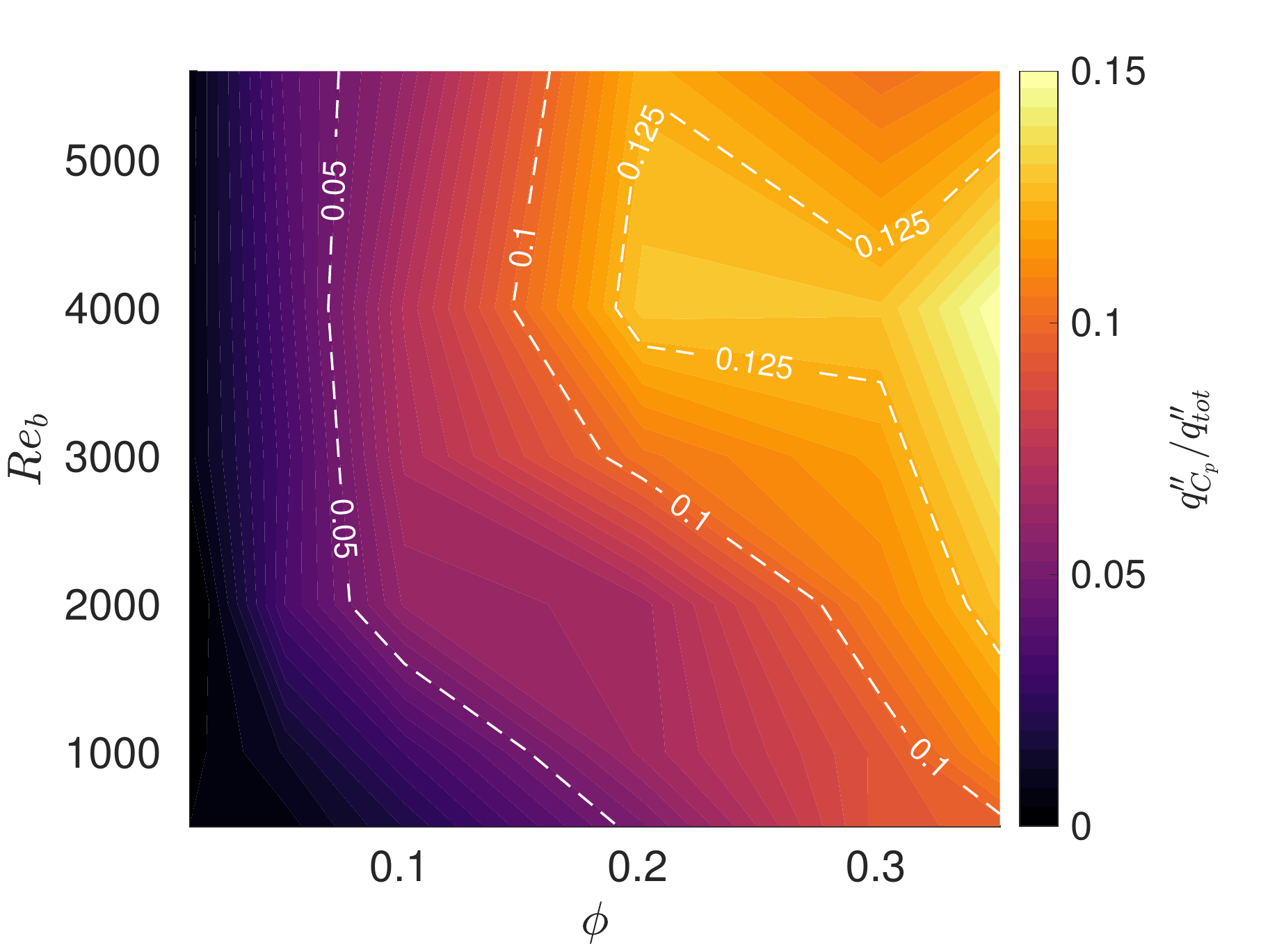}
   \put(-230,160){$(c)$}\\
   \caption{Contour map of the contribution of (a) molecular diffusion, (b) convection by the correlation of the fluid and temperature fluctuations and (c) convection by the correlation of the particle velocity and temperature fluctuations to the total heat transport integrated across the channel for the case with Prandtl number $Pr=7$. The white dashed lines denote different iso-levels as guide for the eye.}
\label{fig:heat_budget_map}
\end{figure}

Figure~\ref{fig:heat_budget_map} shows the contour maps of the relative contribution of (a) molecular diffusion, (b) convection by the fluid and (c) convection by the particle velocity fluctuations to the total heat flux integrated across the channel.
For $Re_b < 2000$ and $\phi < 0.1$, more than $75\%$ of the total heat flux is carried by molecular diffusion.
By increasing the inertia, i.e.\ for $Re_b > 2500$, the share of the convective heat flux exceeds its diffusive counterpart.
The non-monotonic effect of increasing the solid volume fraction on the convection by the fluid velocity fluctuations can be observed from the $75\%$ iso-line shown in figure~\ref{fig:heat_budget_map} (b): the share of $q_{C_f}^{\prime \prime}$ increases when increasing solid volume fraction up to $\phi=0.1$ and decreases beyond that.
The convection is mostly associated with the correlated fluid velocity-temperature fluctuations and the contribution of the correlated particle velocity-temperature fluctuations is at most $15\%$ of the total, which occurs for $Re_b=4000$ and $\phi=35\%$.

\begin{figure}
  \centering
   \includegraphics[width=0.495\textwidth]{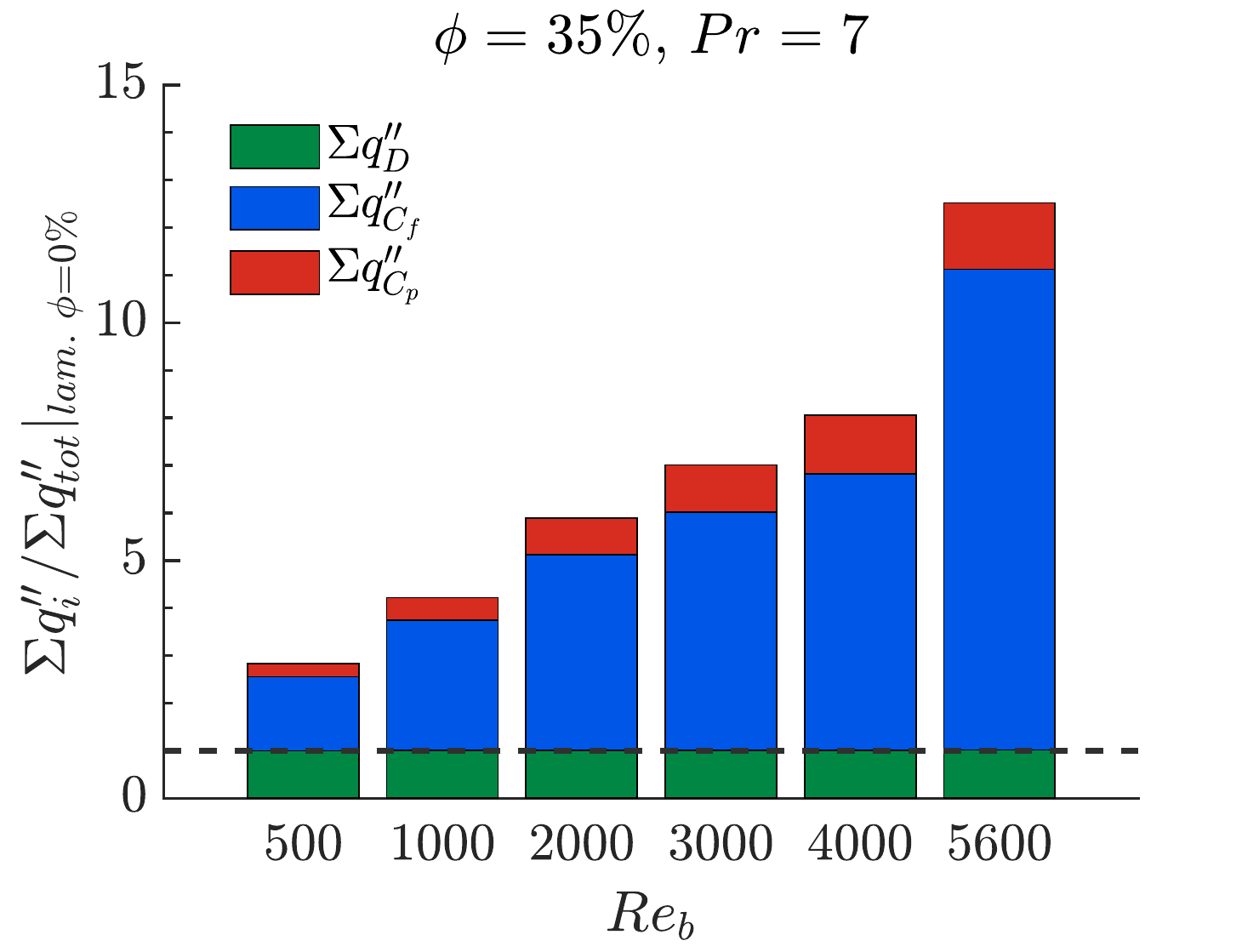}
   \includegraphics[width=0.495\textwidth]{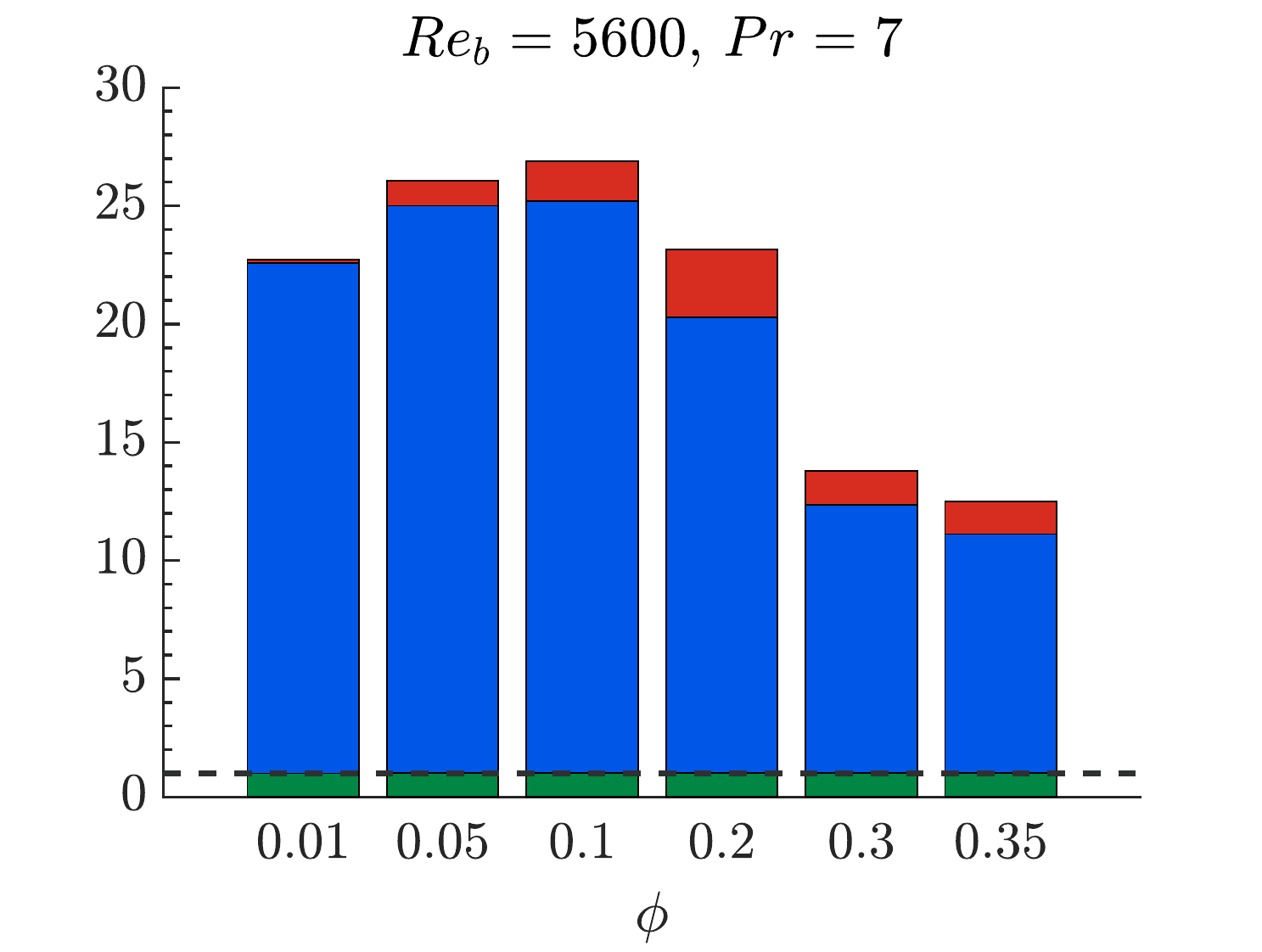}
   \put(-470,160){$(a)$}
   \put(-230,160){$(b)$}\\
   \caption{Wall-normal integral of the different contributions to the total heat flux, normalized by the total heat flux in a laminar single-phase flow for: (a) $\phi=35\%$ and different Reynolds numbers and (b) $Re_b=5600$ and different values of the solid volume fraction $\phi$. The dashed black lines denote the value $\Sigma q_i^{\prime \prime} / \Sigma q^{\prime \prime}_{tot} |_{lam. \, \,\phi=0\%} = 1$.}
\label{fig:heat_budget_bar}
\end{figure}

The integral of the contribution to the total heat flux of each term in Eq.~\eqref{eq:heat_budget1} is depicted in figure~\ref{fig:heat_budget_bar}, where the results are normalized by the total heat flux in a laminar flow in the absence of particles.
Panel (a) of the figure shows the results for different Reynolds numbers at $\phi=35\%$, while the results for different volume fractions at $Re_b=5600$ are depicted  in panel (b).
The dashed line marks the value $\Sigma q_i^{\prime \prime} / \Sigma q^{\prime \prime}_{tot} |_{lam. \, \,\phi=0\%} = 1$ and clearly shows that the integral of the diffusive heat flux across the channel remains constant for the different cases.
The heat flux transferred by convection, on the other hand, increases by increasing inertia (figure~\ref{fig:heat_budget_bar}a) and shows a non-monotonic behavior with increasing the solid volume fraction $\phi$ both for $q_{C_f}^{\prime \prime}$ and $q_{C_p}^{\prime \prime}$; these two contributions first increase and then decrease at higher particle concentrations (figure~\ref{fig:heat_budget_bar}b).
The peak of the heat flux transferred by the fluid-temperature fluctuations occurs at $\phi=10\%$, whereas the largest contribution from the correlated particle velocity-temperature fluctuations is found at $\phi=20\%$.

\begin{figure}
  \centering
   \includegraphics[width=0.495\textwidth]{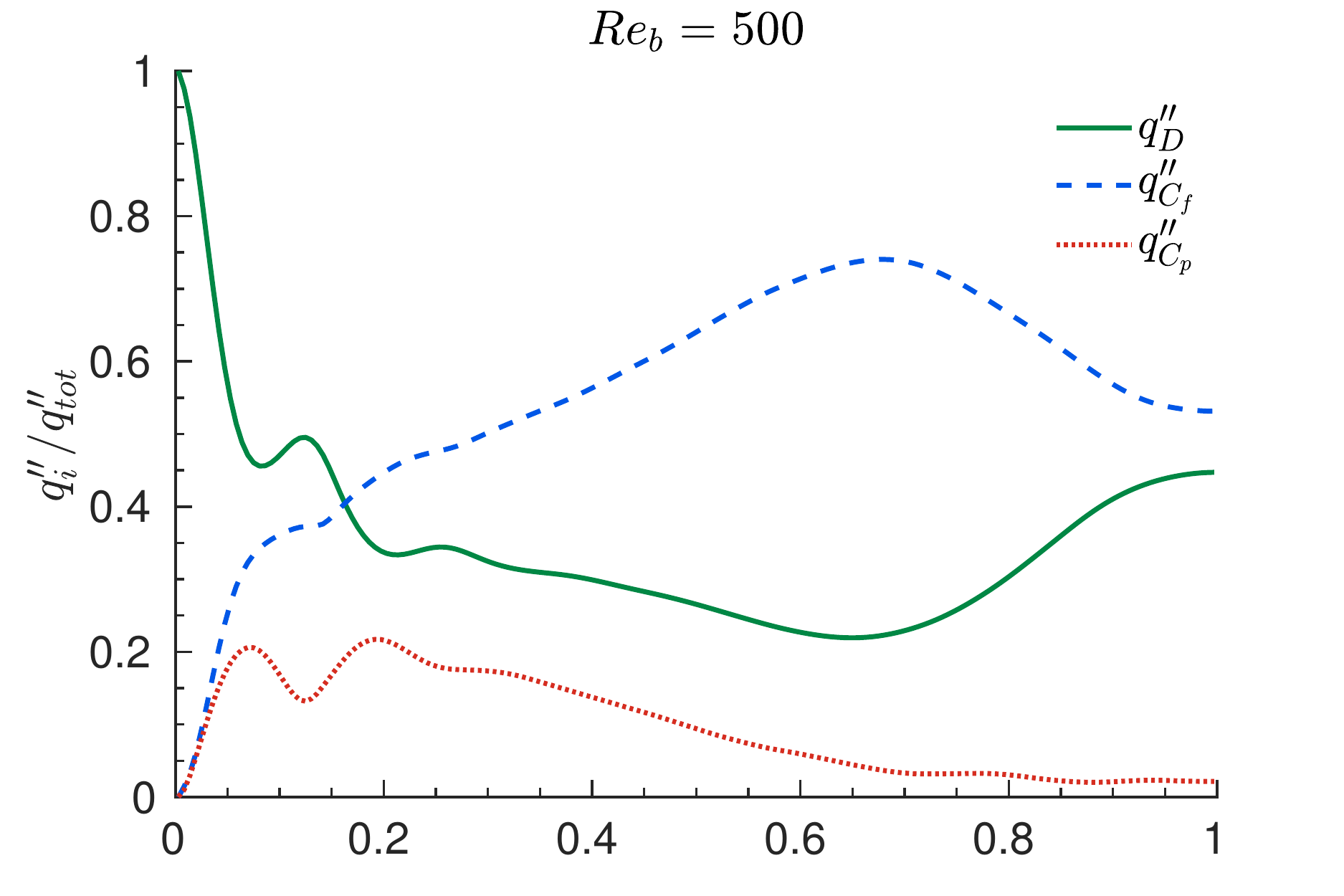}
   \includegraphics[width=0.495\textwidth]{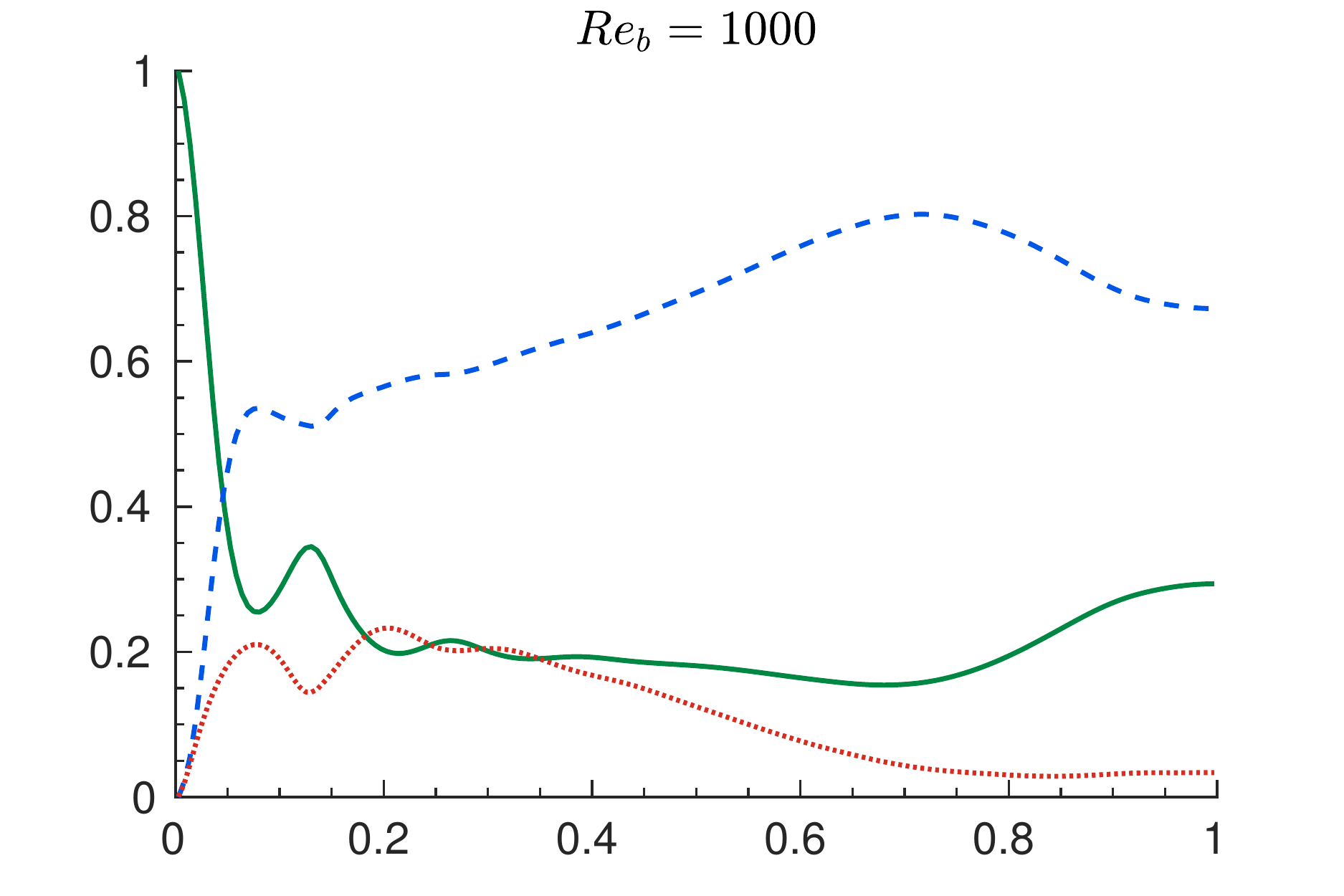}
   \put(-470,150){$(a)$}
   \put(-230,150){$(b)$}\\
   \includegraphics[width=0.495\textwidth]{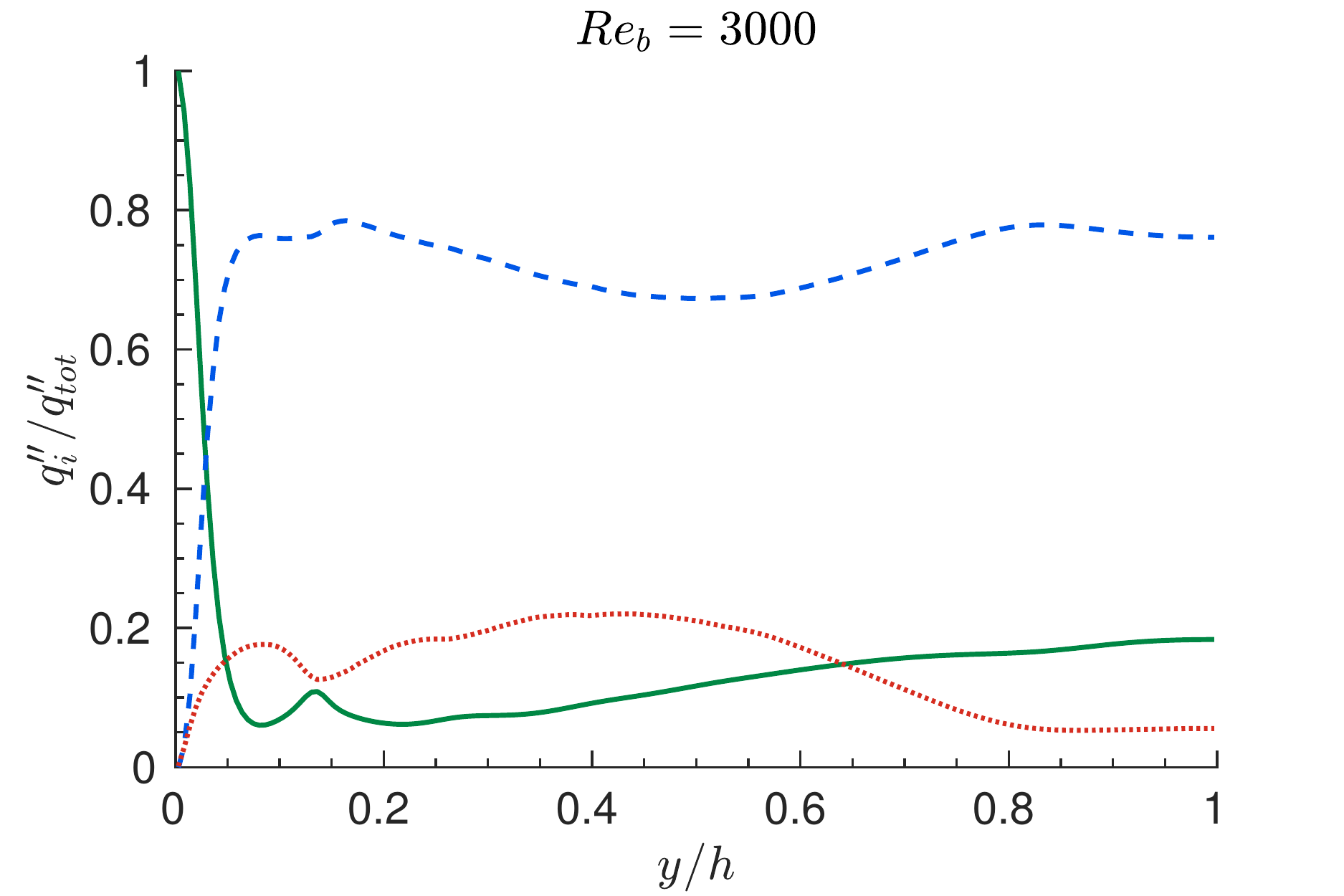}
   \includegraphics[width=0.495\textwidth]{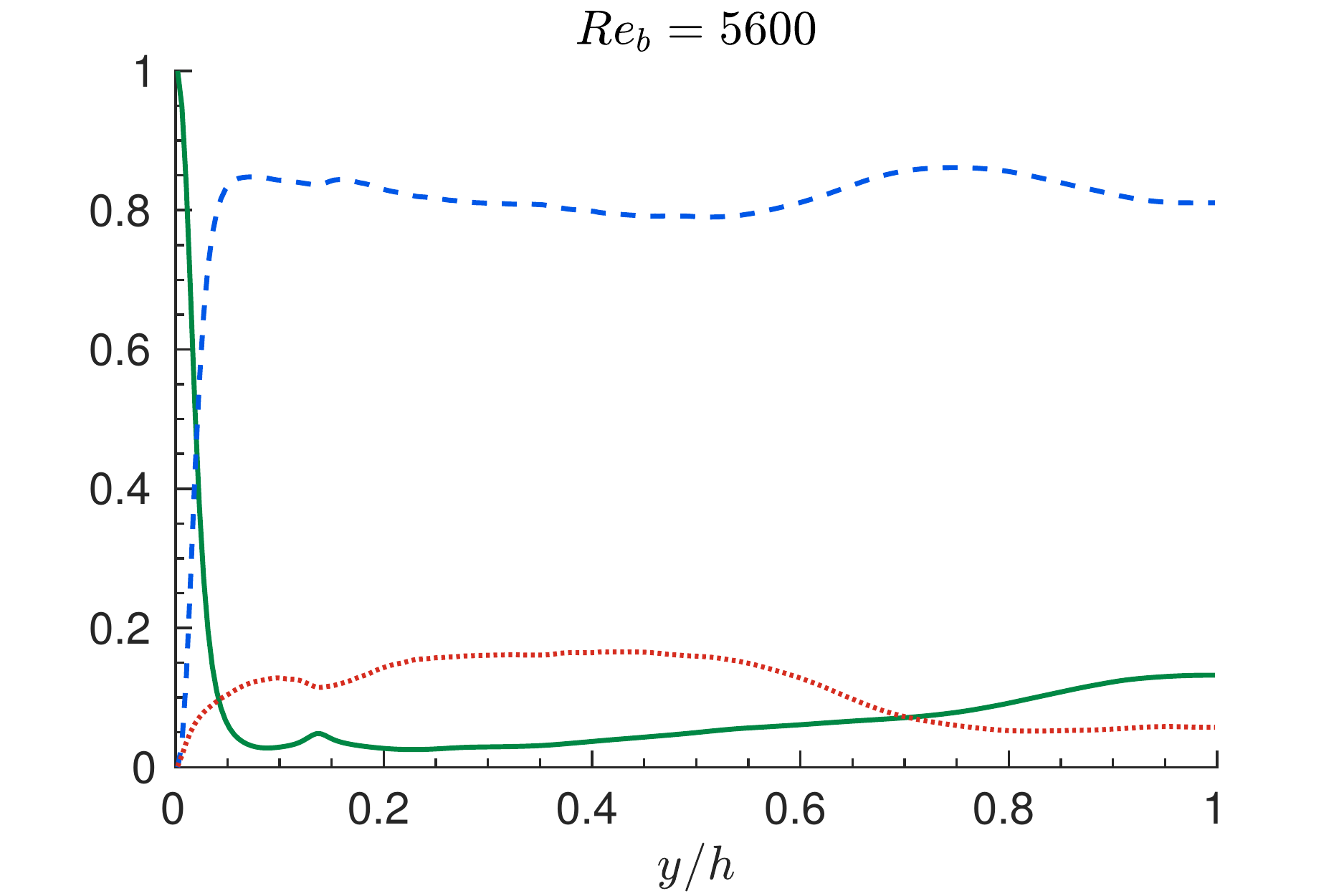}
   \put(-470,150){$(c)$}
   \put(-230,150){$(d)$}\\
   \caption{Heat flux budget wall-normal profiles for the volume fraction $\phi=35\%$ and $Pr=7$: (a) $Re_b=500$, (b) $Re_b=1000$, (c) $Re_b=3000$ and (d) $Re_b=5600$.}
\label{fig:heat_profile_Re}
\end{figure}

To document how the contribution of each transport mechanism changes across the channel, we present the wall-normal profiles of the different terms in equation (\ref{eq:heat_budget1}) for a number of selected cases.
In particular, figure~\ref{fig:heat_profile_Re} displays the profiles at the volume fraction $\phi=35\%$ and different Reynolds numbers.
At $Re_b=500$, when the flow is laminar, the diffusion is the main mechanism of the heat transfer close to the wall, $y/h < 0.15$.
Further away from the wall, the convection by correlated fluid velocity fluctuations $q_{C_f}^{\prime \prime}$ takes over and reaches a peak around $y/h=0.7$, before reducing approaching the centerline.
Convection by the particle velocity fluctuations $q_{C_p}^{\prime \prime}$ on the other hand, has two consecutive peaks close to the wall ($y/h<0.2$) and vanishes in the core region, following the particle distribution.

When increasing the Reynolds number, the distance at which the heat transfer term $q_{C_f}^{\prime \prime}$ exceeds $q_{D}^{\prime \prime}$ moves closer to the wall. 
Also, the profiles of heat transfer by the velocity fluctuations become more uniform across the channel in both phases, although $q_{C_p}^{\prime \prime}$ vanishes in the core region even for $Re_b=5600$.

\begin{figure}
  \centering
   \includegraphics[width=0.495\textwidth]{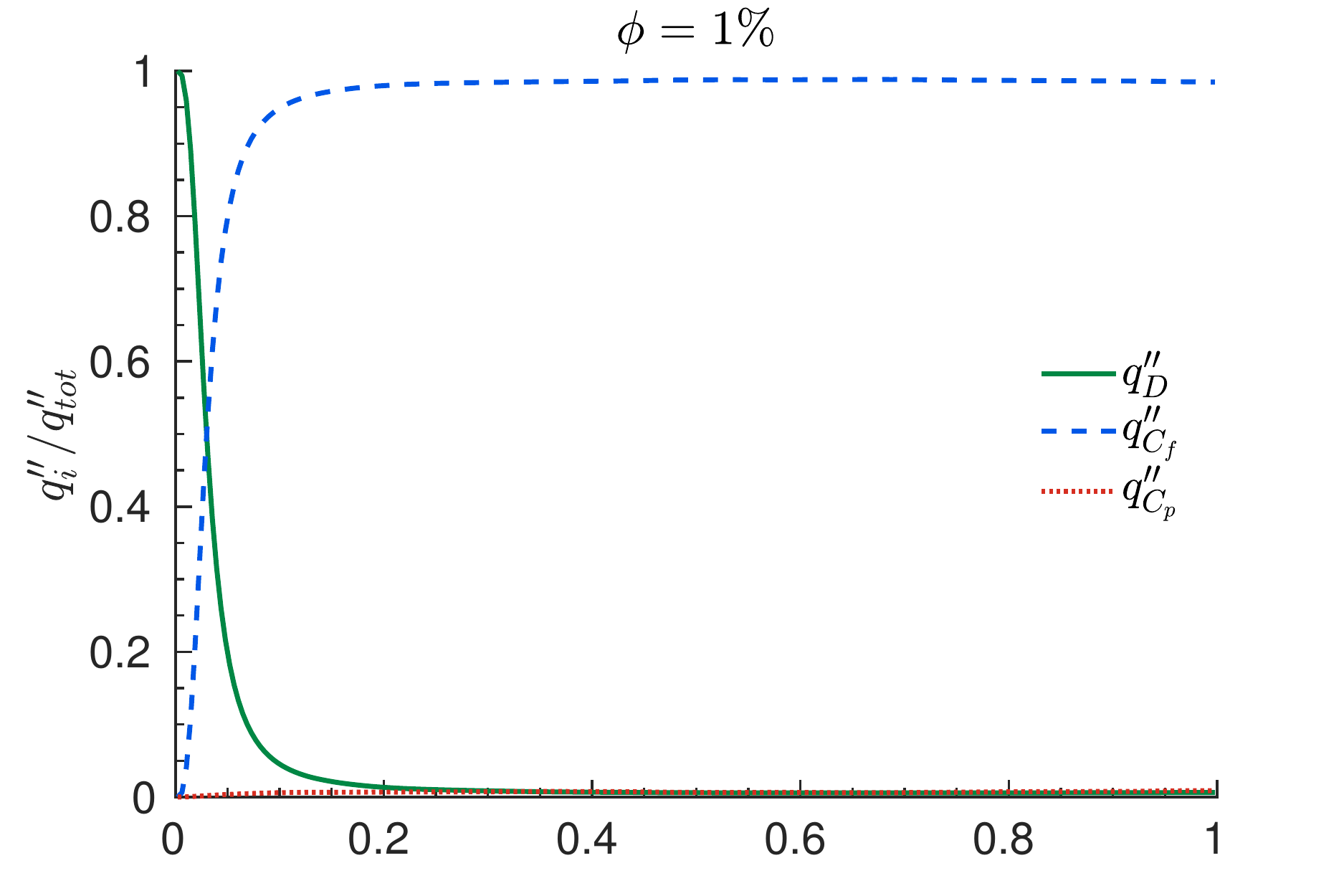}
   \includegraphics[width=0.495\textwidth]{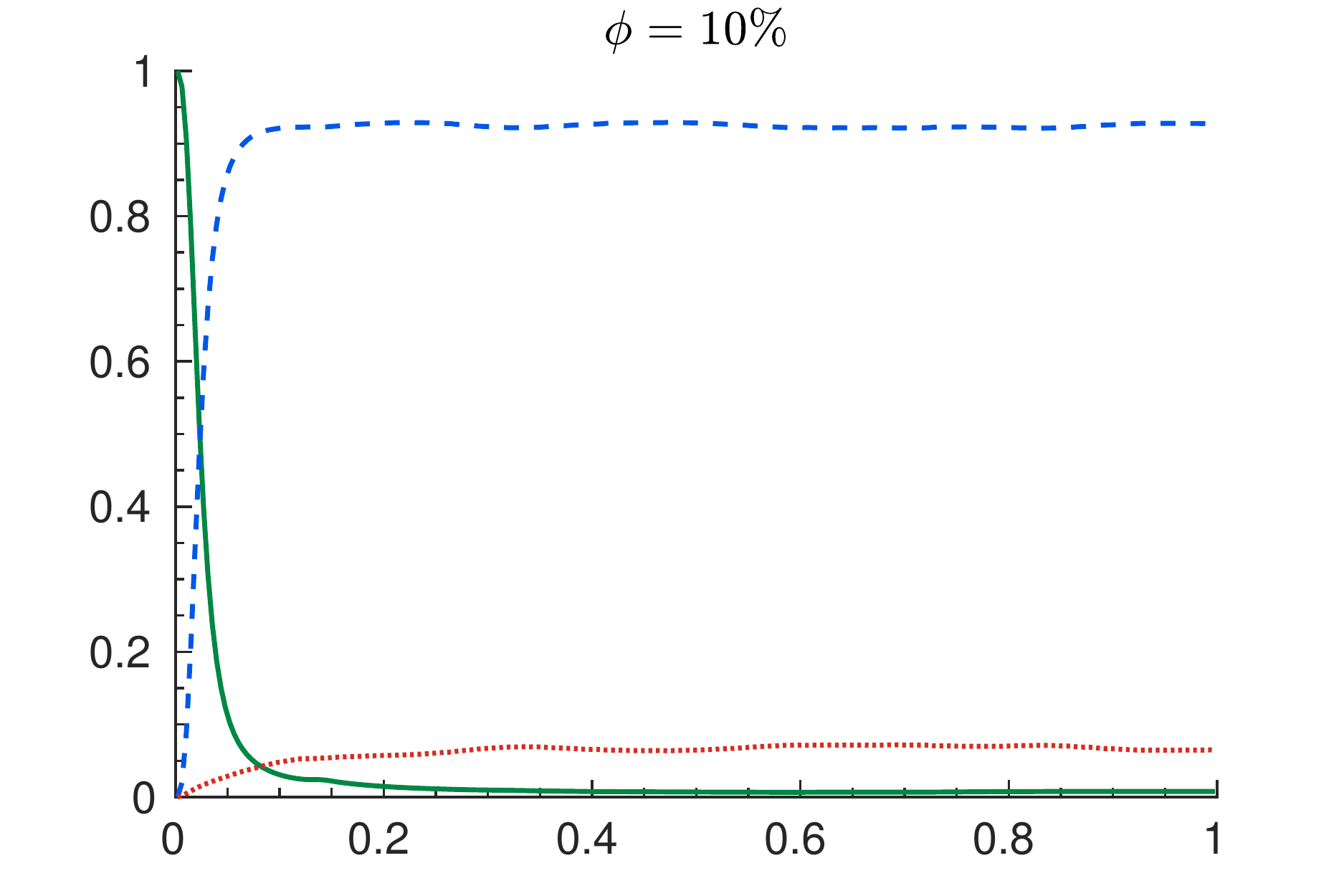}
   \put(-470,150){$(a)$}
   \put(-230,150){$(b)$}\\
   \includegraphics[width=0.495\textwidth]{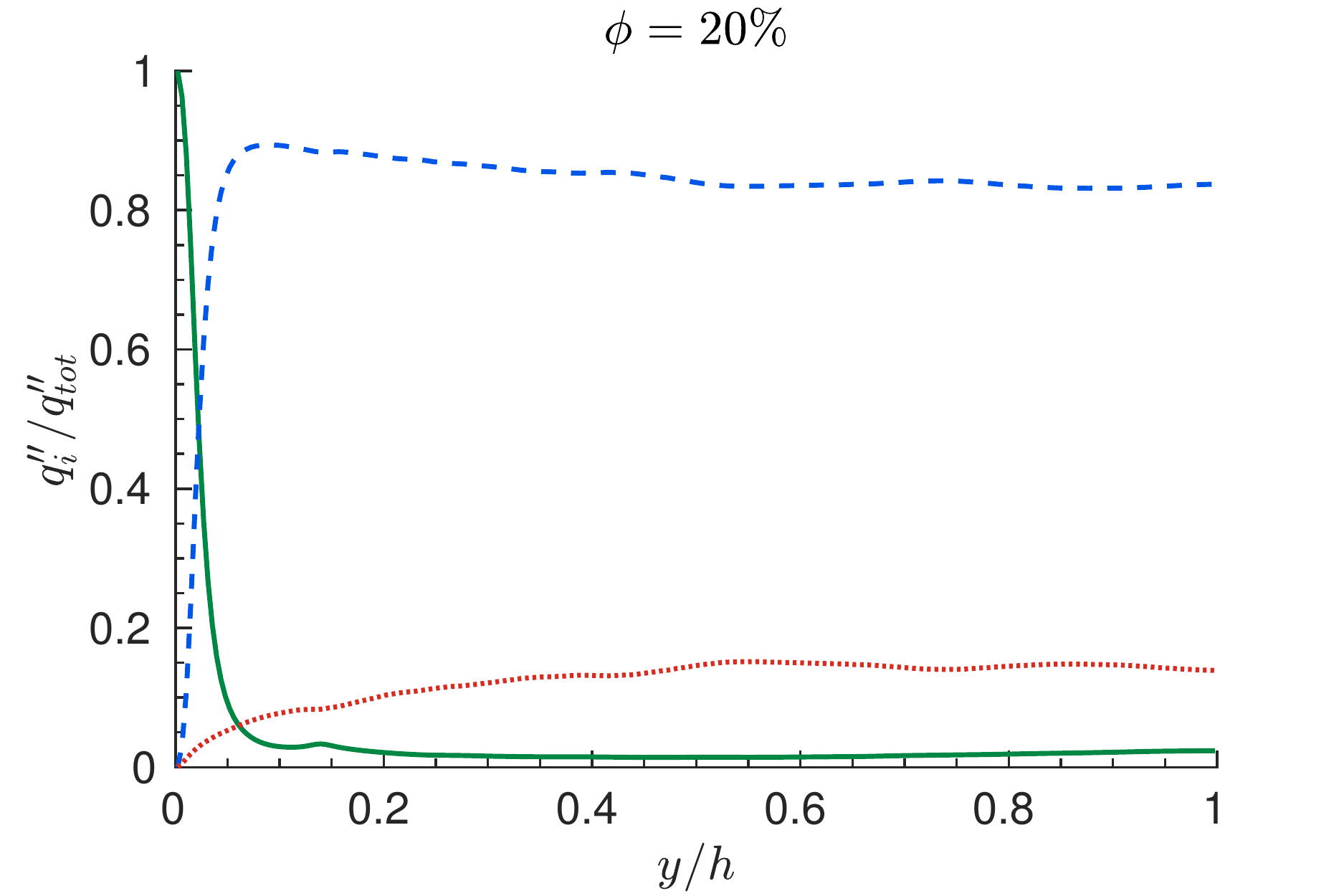}
   \includegraphics[width=0.495\textwidth]{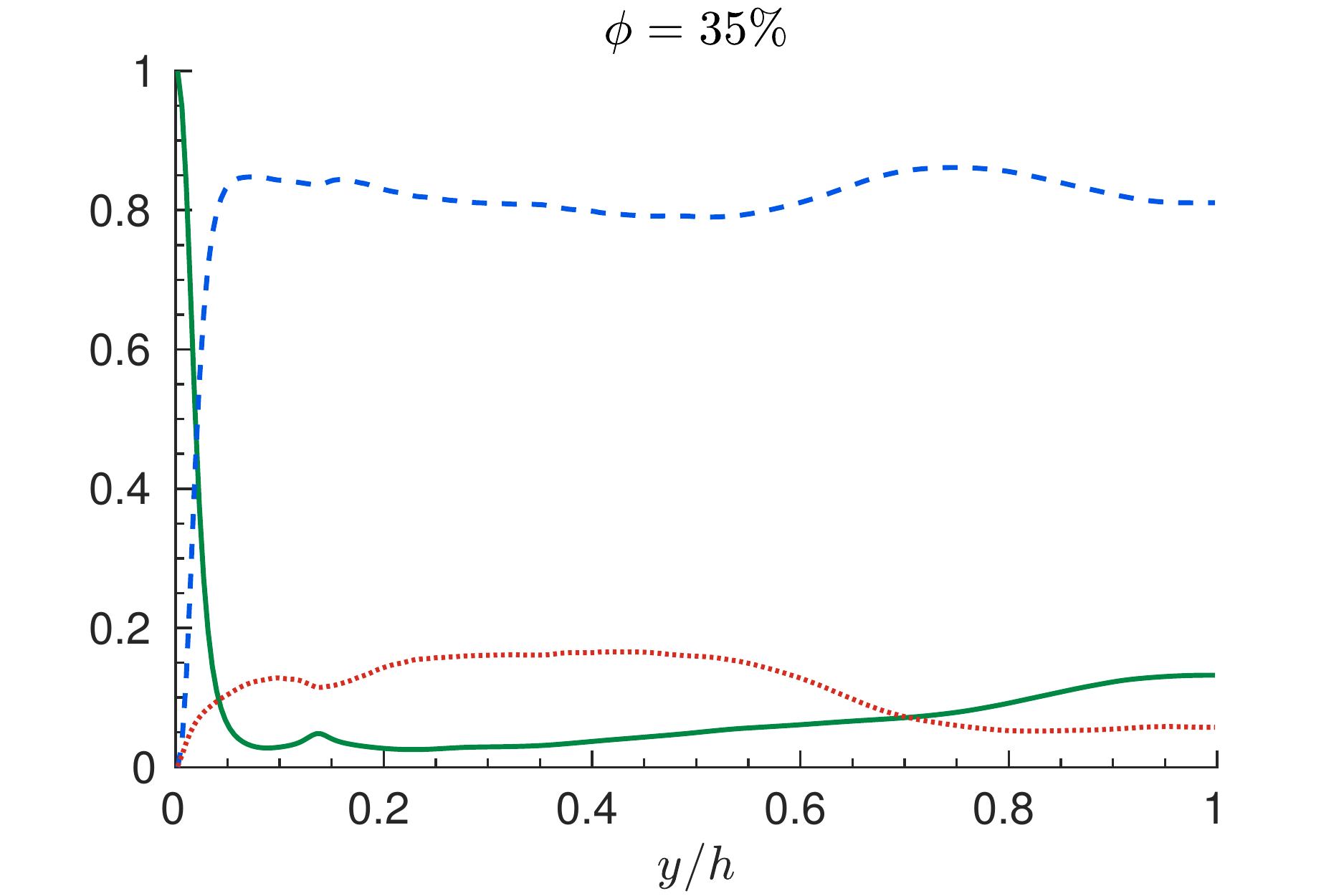}
   \put(-470,150){$(c)$}
   \put(-230,150){$(d)$}\\
   \caption{Heat flux budget wall-normal profiles for the Reynolds number $Re_b=5600$ and $Pr=7$: (a) $\phi=1\%$, (b) $\phi=10\%$, (c) $\phi=20\%$ and (d) $\phi=35\%$.}
\label{fig:heat_profile_phi}
\end{figure}

To focus on the effect of the particle volume fraction, we show the wall-normal profiles of the terms appearing int he heat-transfer budget at $Re_b=5600$ and for different volume fractions $\phi=1\%\mbox{-}35\%$ in figure~\ref{fig:heat_profile_phi}.
At low concentrations, the heat flux is almost completely due to the convection by the fluid velocity fluctuations.
By increasing the solid volume fraction, the share of $q_{C_p}^{\prime \prime}$ increases almost uniformly across the channel and reaches the maximum integral contribution at $\phi=20\%$, as shown in figure~\ref{fig:heat_budget_bar} (b).
Finally, as discussed above, the contribution to the total heat flux by the particle motion reduces close to the centerline at the higher solid volume fractions.

\subsection{Flow properties and particle dynamics}

\begin{figure}
  \centering
   \includegraphics[width=0.495\textwidth]{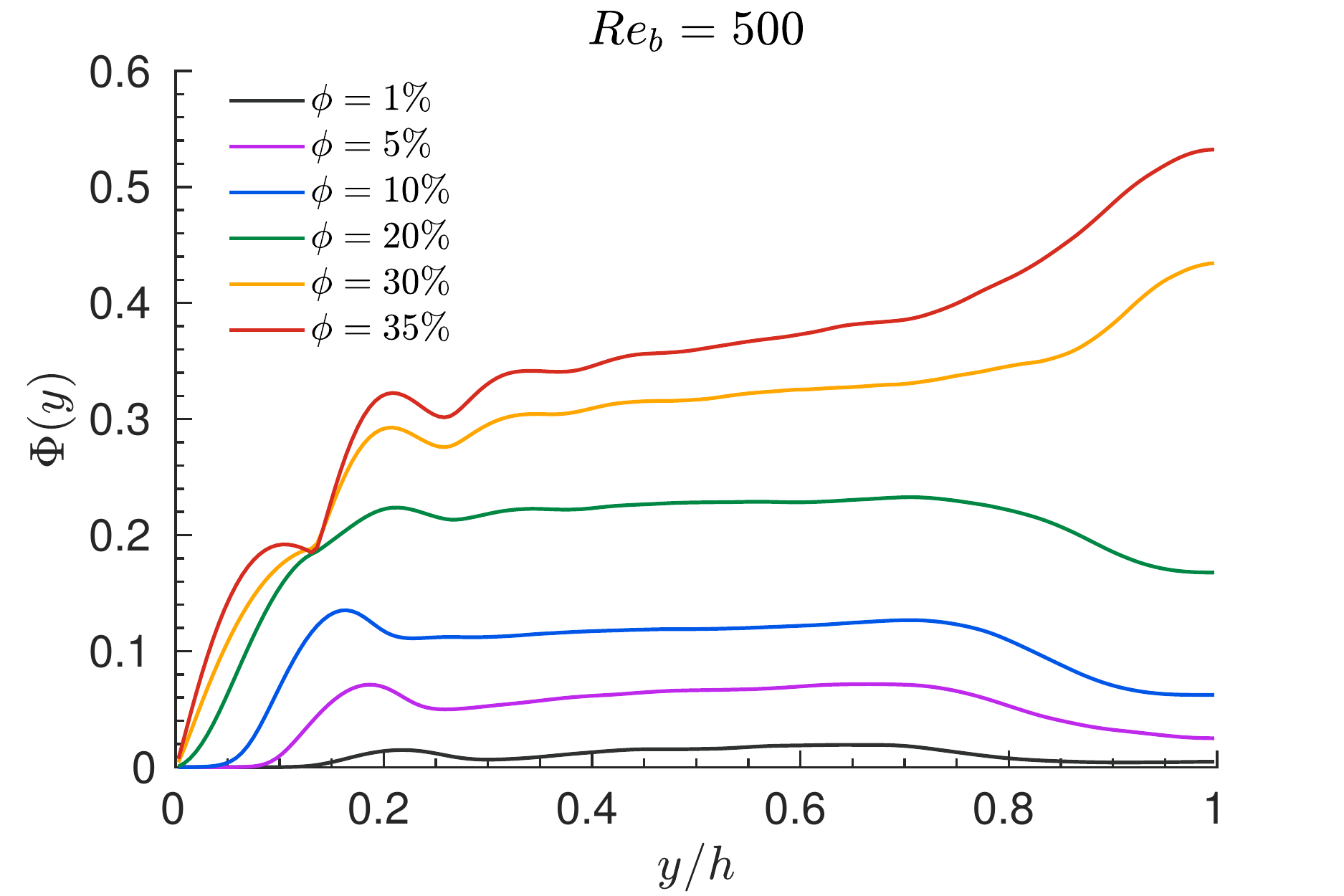}
   \includegraphics[width=0.495\textwidth]{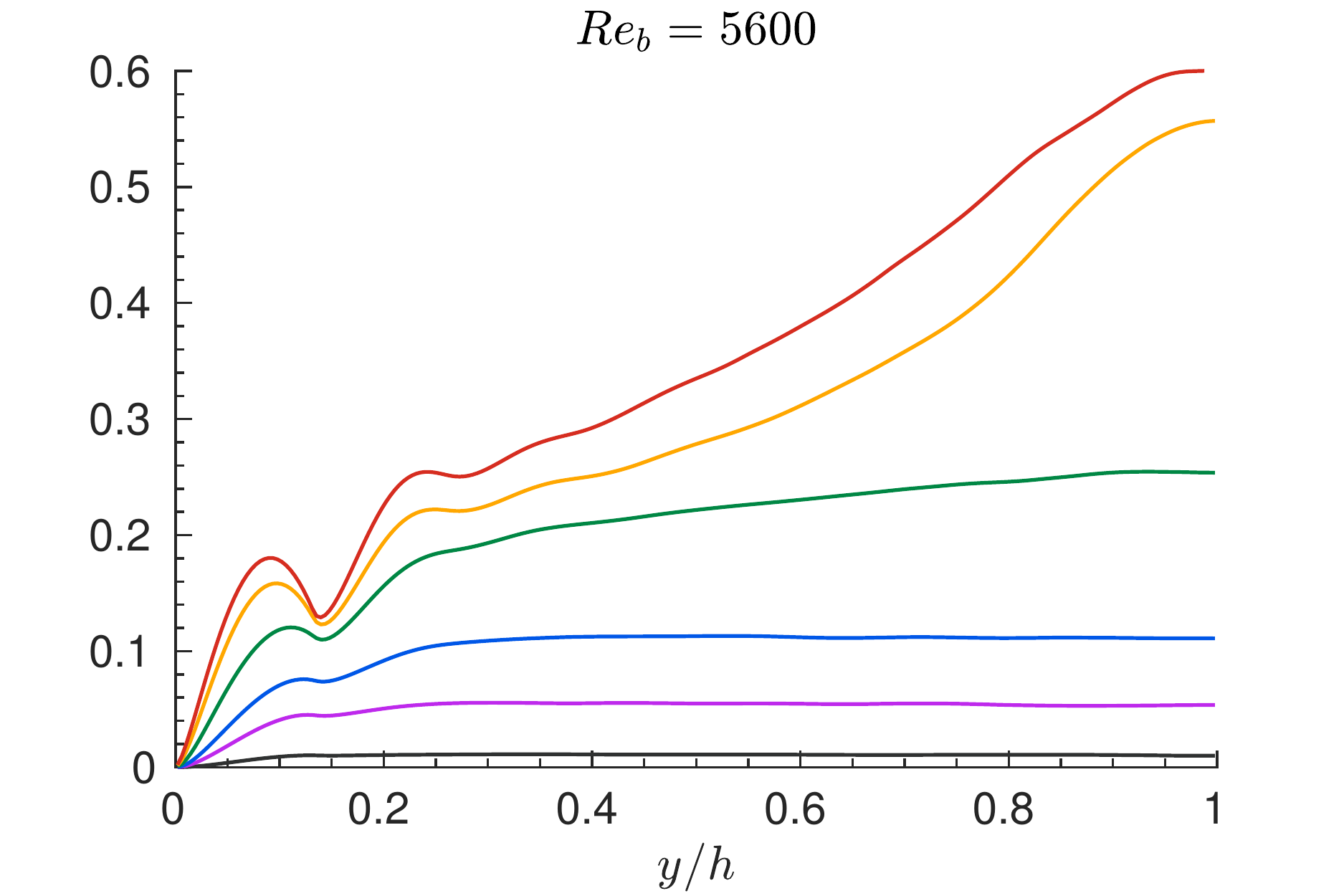}
   \put(-470,150){$(a)$}
   \put(-230,150){$(b)$}\\
   \caption{Wall-normal profiles of the local volume fraction $\Phi (y)$ for: (a) $Re_b=500$ and (b) $Re_b=5600$ for the solid volume fractions indicated in the legend.}
\label{fig:local_concentration}
\end{figure}

In this section, we focus on the mean statistics of the fluid and particle phases to understand the mechanisms behind the heat transfer enhancement reported in the previous section.
The wall-normal profiles of the local particle volume fraction $\Phi(y)$ are shown in figure~\ref{fig:local_concentration} for different Reynolds numbers and nominal volume fractions.
In the laminar regime, see panel (a), at low to moderate volume fractions $1\% \leq \phi \leq  20\%$,  the particles accumulate in the intermediate region between the wall and the channel centerline, $0.2 \leq y/h \leq 0.8$.
This is reminiscent of the Segre-Silberberg effect \citep{Segre1961}, an inertial effect resulting from the balance between the Saffman lift \citep{saffman-JFM-1965}, inhomogeneous shear rate and wall effects; see also \cite{yeo-JFM-2010}.
In the dilute turbulent regime ($\phi < 20\%$), on the other hand, the local volume fraction is almost uniform across the channel, except close to the wall due to excluded volume effects, see panel (b) of the same figure. In this case, turbulent mixing induces a homogeneous solid concentration in the bulk, whereas particles experience asymmetric interactions with the walls.

Increasing the volume fraction, $\phi \geq 30\%$, we observe a significant accumulation of particles in the core region, with maximum at the channel centreline.
This behavior is similar to the shear-induced migration observed in laminar inhomogeneous shear flows, where it is explained by particle-particle interactions and the imbalance of the normal stresses in the wall-normal direction  \cite[see among others][]{nott-JFM-1994,guazzelli-book-2011}.
Here, we observe it in the presence of inertia
both at  $Re_b=500$ and $5600$. Comparing the two panels of the figure, we see that the concentration at the centreline is largest for the highest Reynolds number considered, indicating that the migration increases when increasing the inertia. Furthermore, the results in \cite{NiaziArdekani} show that migration is more evident for larger particles at fixed volume fraction and Reynolds number, and those in \cite{Fornari2016} that migration is evident already at $\Phi=5\%$ when the particle density is 10 times the fluid density, confirming the role of particle inertia for the shear-induced migration towards the center line reported here. 

The local volume fraction at the core region reaches values $\Phi (y) > 50\%$ and the particles approach a random loose-packing configuration. 
Note also the local maxima of the local volume fraction around $y/h = 0.1$, corresponding to the particle layering in the vicinity of the wall, see \cite{Costa2016}. Once a particle approaches the wall it tends to stay there, since the interaction with the neighboring suspended particles is asymmetric and the strong near-wall lubrication force hinders departing motions \cite[see also][]{lashgari-IJMF-2016}.

\begin{figure}
  \centering
   \includegraphics[width=0.495\textwidth]{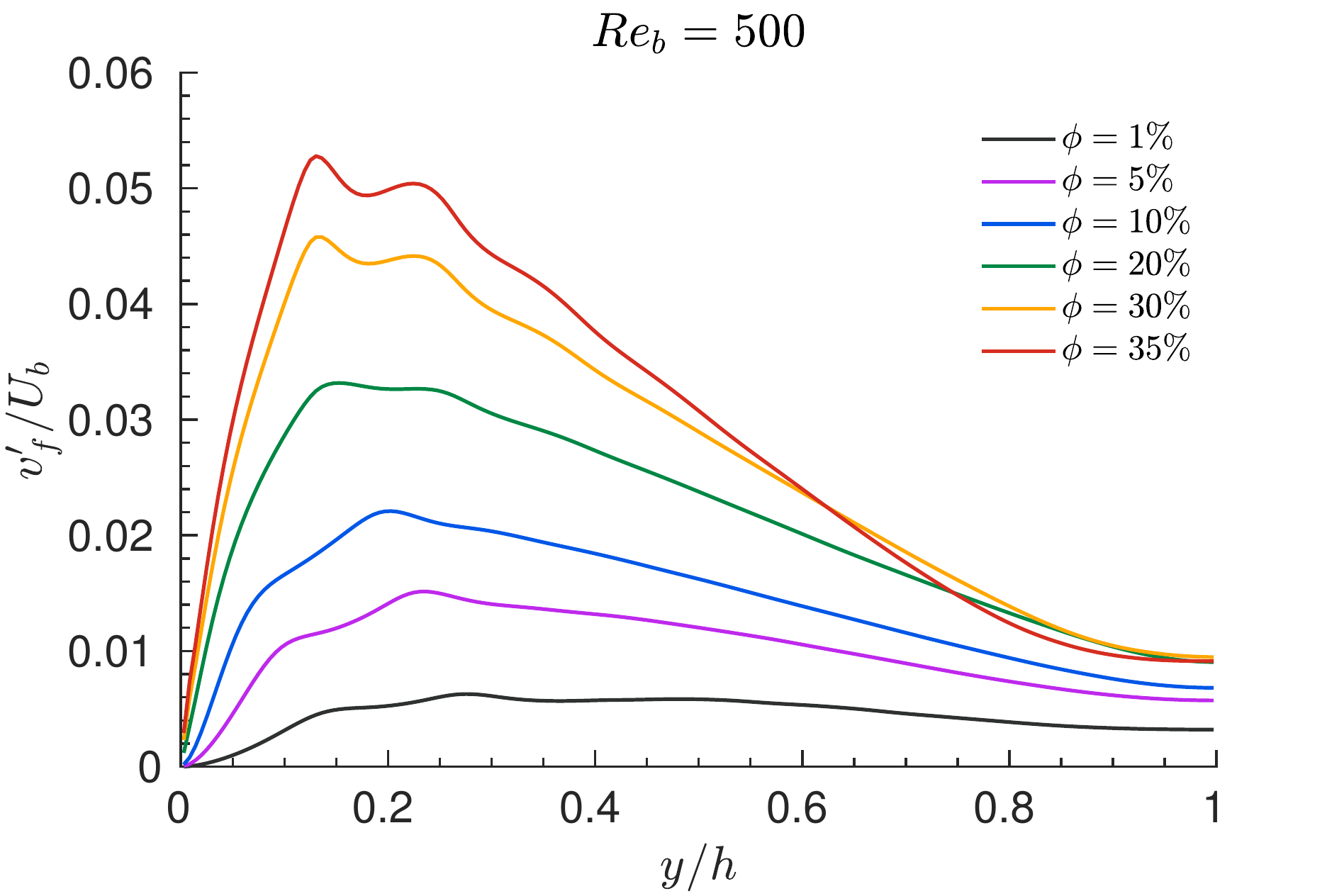}
   \includegraphics[width=0.495\textwidth]{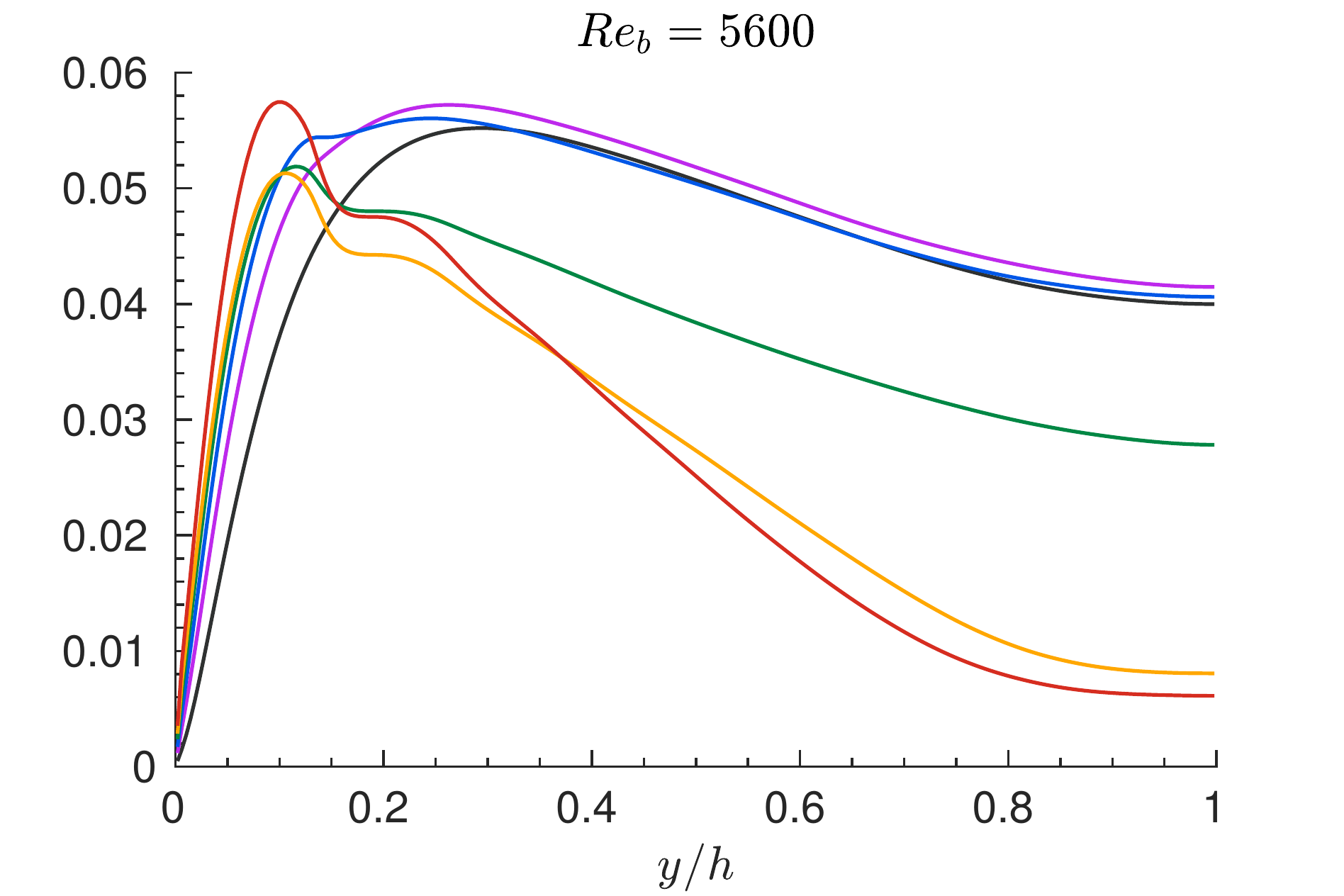}
   \put(-470,150){$(a)$}
   \put(-233,150){$(b)$}\\
   \caption{Wall-normal profiles of the r.m.s.\ of the fluid wall-normal velocity fluctuations, normalized by the bulk fluid velocity $v_f^{\prime} / U_b$ at: (a) $Re_b=500$ and (b) $Re_b=5600$. }
\label{fig:wall_normal_vel_fluid}
\end{figure}

Figure~\ref{fig:wall_normal_vel_fluid} displays the root-mean-square (r.m.s.) of the wall-normal velocity fluctuations of the fluid phase for the cases with $Re_b=500$ in panel (a) and $5600$ in panel (b). 
At $Re_b=500$, we note a peak in the wall-normal velocity profile close to the wall, which intensifies significantly at high concentrations. 
This peak can be induced by particles approaching and departing from the wall whose local concentration increases with the volume fraction, see figure~\ref{fig:local_concentration}(a) at $y/h \approx 0.15$.
Note also that for all the investigated volume fractions, the r.m.s.\ of the wall-normal velocity fluctuation rapidly decays in the core region. 

In the turbulent regime ($Re_b=5600$), see panel (b) of figure~\ref{fig:wall_normal_vel_fluid}, the fluid wall-normal velocity fluctuations are reminiscent of the single-phase flow for solid concentrations $\phi \leq 10\%$, i.e.\ we see an almost constant distribution across the channel with a small peak close to the wall, where the addition of the solid phase 
slightly increases the velocity fluctuations. At high volume fractions though, $\phi \geq 30\%$, the profiles become similar to those at $Re_b=500$, with a peak close to the wall, characterising increasing turbulent activity associated to the particle wall layer, and vanishing fluctuations at the centreline, where particles tend to accumulate in a random loosely packed configuration. 
The flow at $\phi=20\%$ and $Re_b=5600$ displays an intermediate behavior, i.e.\ we note a peak close to the wall, and not negligible, yet attenuated,  fluctuations in the core region.

\begin{figure}
  \centering
   \includegraphics[width=0.495\textwidth]{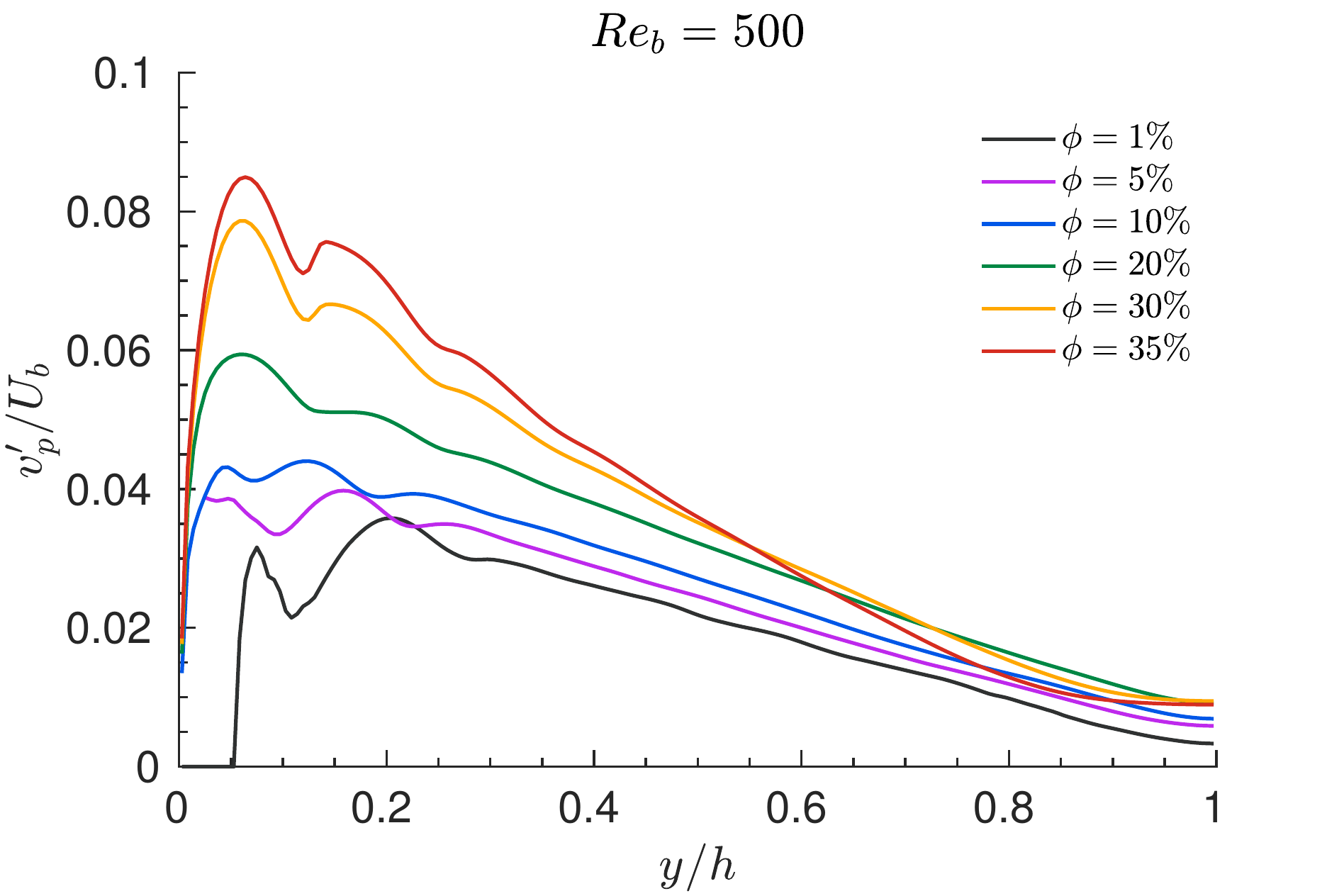}
   \includegraphics[width=0.495\textwidth]{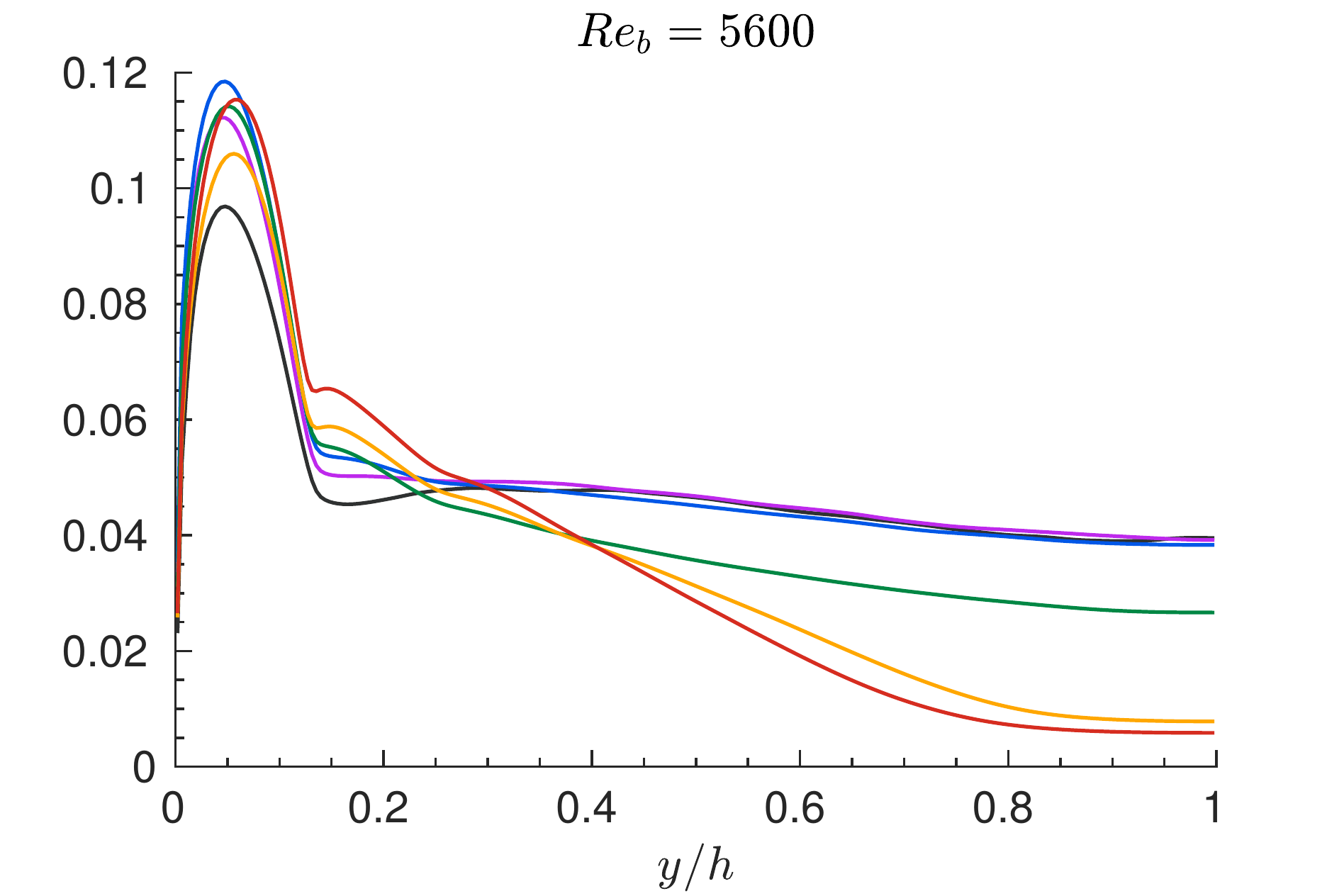}
   \put(-470,150){$(a)$}
   \put(-233,150){$(b)$}\\
   \caption{Wall-normal profiles of the r.m.s. of the particle wall-normal velocity fluctuations, normalized by the bulk fluid velocity $v_p^{\prime} / U_b$ at: (a) $Re_b=500$ and (b) $Re_b=5600$.}
\label{fig:wall_normal_vel_particle}
\end{figure}

The wall-normal profiles of the r.m.s.\ of the particle wall-normal velocity fluctuations are depicted in figure~\ref{fig:wall_normal_vel_particle}.
In the laminar regime (panel a) the profiles are similar to those of the fluid phase, see figure~\ref{fig:wall_normal_vel_fluid} (a). The main difference is that the near-wall peak 
in the dilute regime is more evident for the particle phase than for the fluid phase.
This peak is most likely due to the particle collisions with the wall 
so that the particle wall-normal velocity fluctuations can be considered as the cause of the near-wall peak in the fluid phase fluctuation profiles. 
This effect is, expectedly, proportional to the solid phase concentration and is therefore less significant in the dilute regime, see figure~\ref{fig:wall_normal_vel_fluid} (a).
 
 In the turbulent regime, (panel b of figure~\ref{fig:wall_normal_vel_particle}), we observe a very distinct peak in the particle wall-normal velocity fluctuation profiles in the region $y/h \approx 0.05$. Moving away from the wall,  the level of the velocity fluctuations remain uniform across the channel at lower volume fractions, $\phi \leq 10\%$, to a value $v_p^{\prime} / U_ b \approx 0.05$. 
 More interestingly, we observe significantly smaller values of the wall-normal particle velocities close to the centerline ($y/h > 0.4$) at the highest volume fractions, $\phi \geq 30\%$, with values 
 similar to those of the laminar cases. This attenuation reflects the significant particle accumulation discussed above, and explains the reduced heat transfer in turbulent flows at high particle volume fractions.  Indeed, the flow near the centreline is characterised by a high concentration of particles, moving almost as a compact aggregate, with reduced mixing and heat transfer therefore dominated by molecular diffusion.

\begin{figure}
  \centering
   \includegraphics[width=0.495\textwidth]{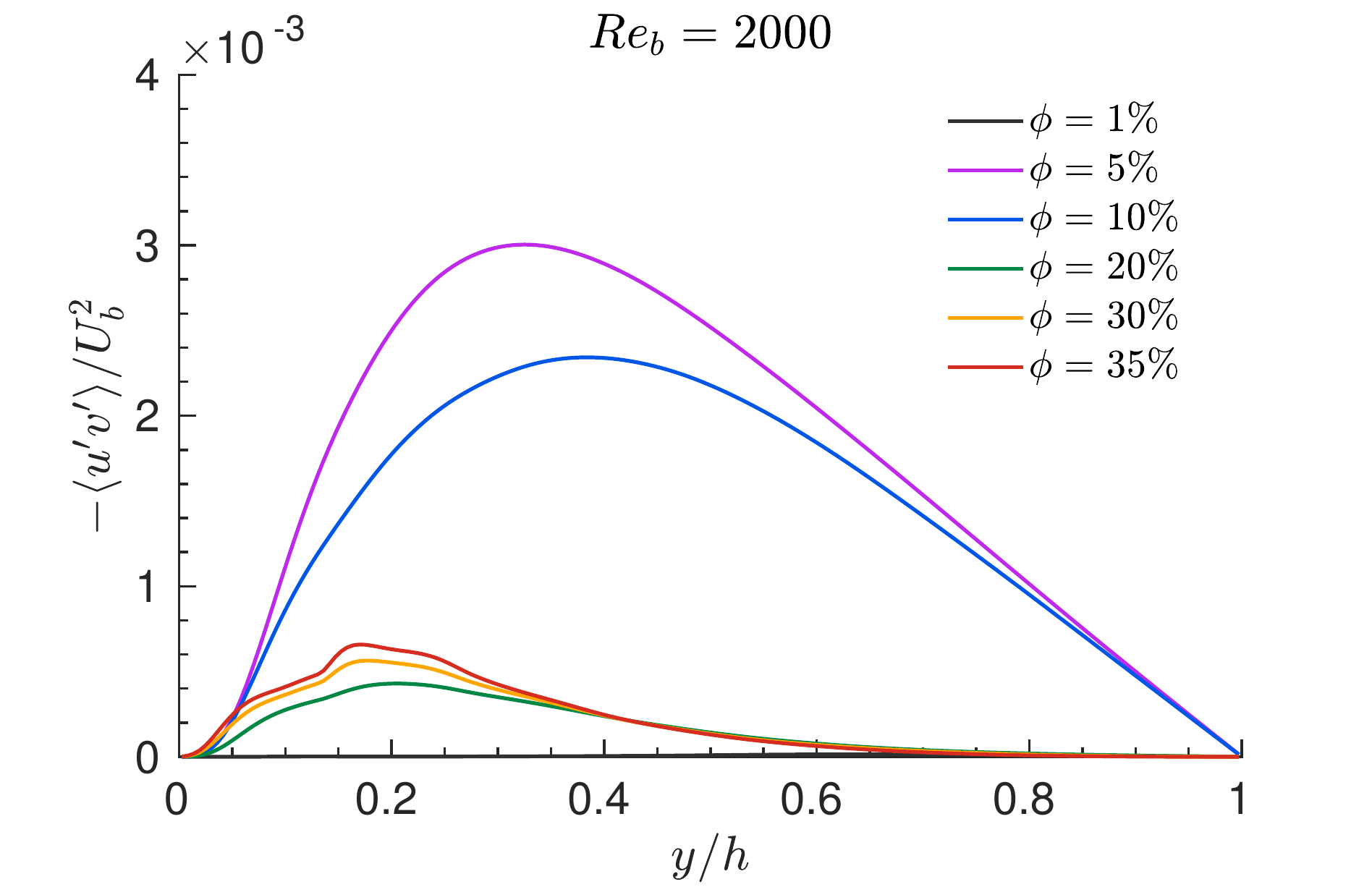}
   \includegraphics[width=0.495\textwidth]{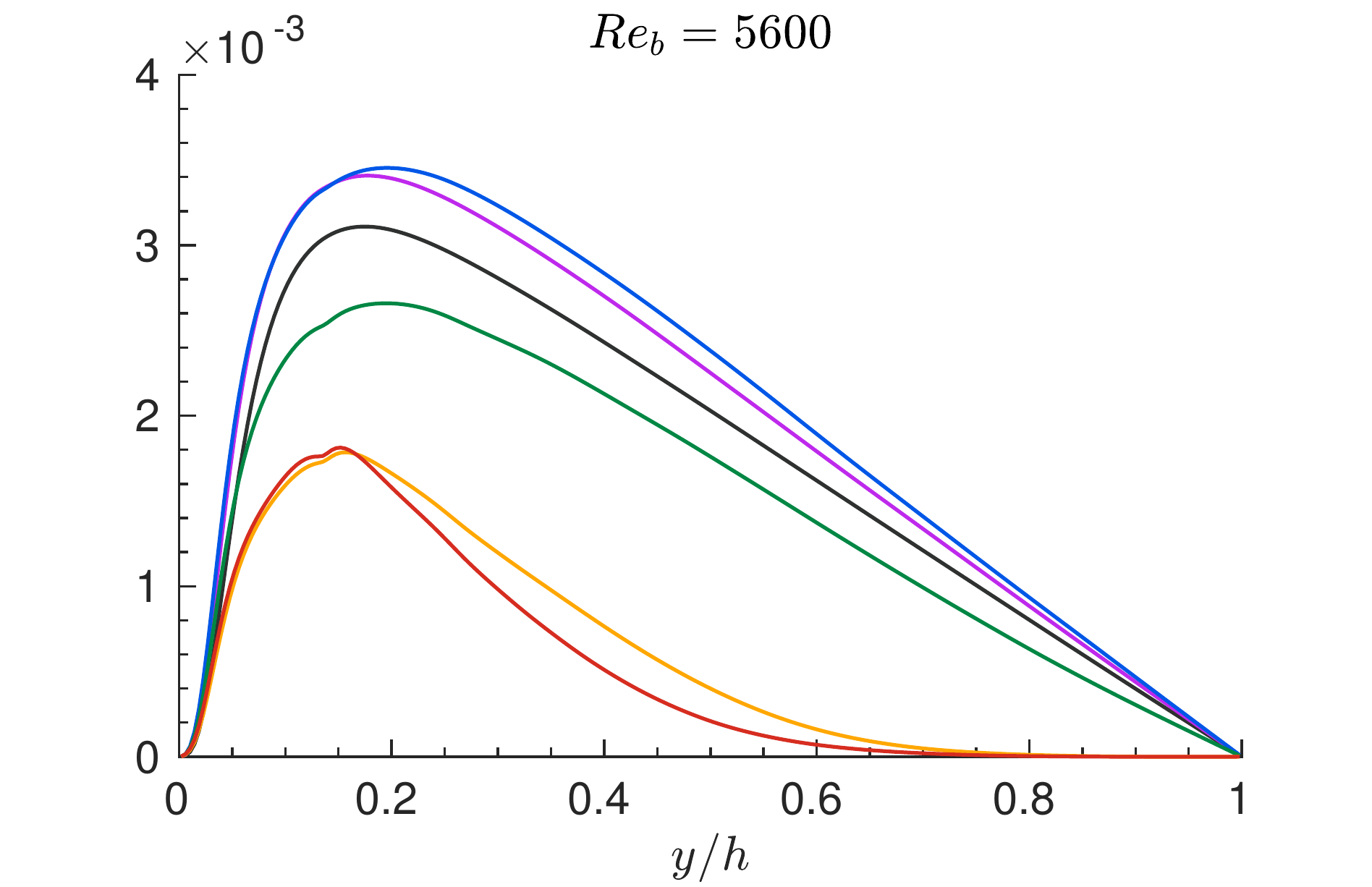}
   \put(-470,150){$(a)$}
   \put(-230,150){$(b)$}\\
   \caption{Wall-normal profiles of the Reynolds shear stress, normalized by the bulk fluid velocity $-\langle u^{\prime} v^{\prime} \rangle / U_b^2$ at: (a) $Re_b=2000$  and (b) $Re_b=5600$.}
   \label{fig:RSS}
\end{figure}

To quantify the effect of the suspended phase on the turbulent activity-- in general, an increase close to the wall and decrease in the bulk at high $\Phi$, we present the wall-normal profiles of the Reynolds shear stress $(- \langle u^{\prime} v^{\prime} \rangle)$ for the different volume fraction considered and $Re_b = 2000$ and $5600$ in figure~\ref{fig:RSS}. The results clearly show the non-monotonic behavior which was observed for the heat transfer efficiency too.
At $Re_b=2000$, the flow remains laminar at $\phi=1\%$, while the noise introduced by the presence of the particles triggers a turbulent-like behavior at $5\% \leq \phi \leq 10\%$; further increasing the volume fraction, flow becomes laminar-like again due to increased dissipation.
At $Re_b=5600$, the Reynolds shear stress increases across the channel for particle volume fractions up to $\phi=10\%$, while it decreases for $\phi \geq 20\%$. Interestingly, for the flows with $\phi \geq 30\%$ the turbulent activity decreases significantly close to the wall and vanishes in the core region, $y/h > 0.7$, which clearly shows that at this solid concentrations the flow consists of a laminar-like core packed with particles moving as a plug.

To conclude, we would like to take this opportunity the complete and extend the analysis previously presented in \cite{Lashgari-PRL-2014} and \cite{lashgari-IJMF-2016} on the different flow regimes in particulate channel flows and relate those to the observations here on the heat transfer in the suspensions.

\begin{figure}
  \centering
   \includegraphics[width=0.495\textwidth]{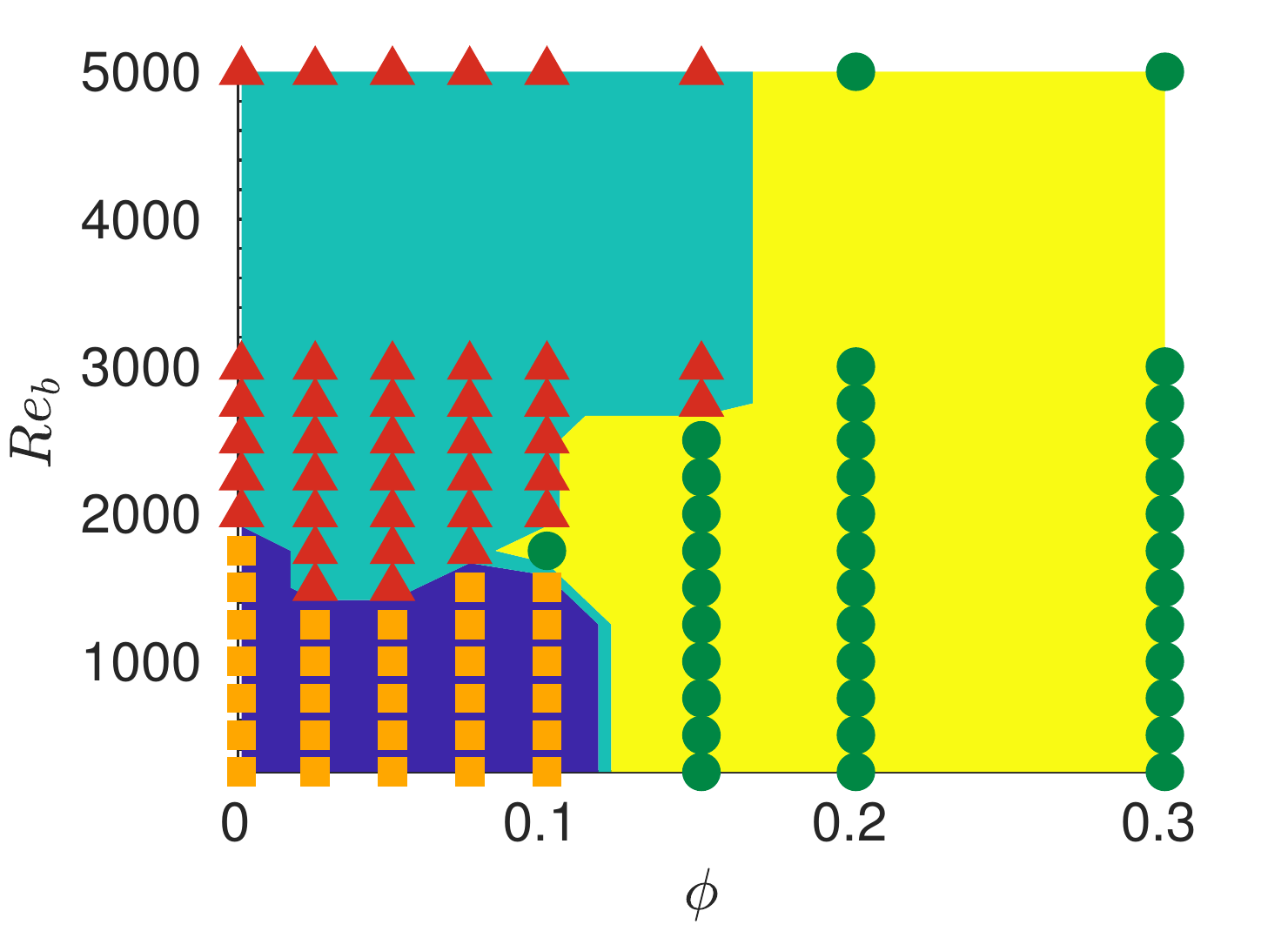}
   \includegraphics[width=0.495\textwidth]{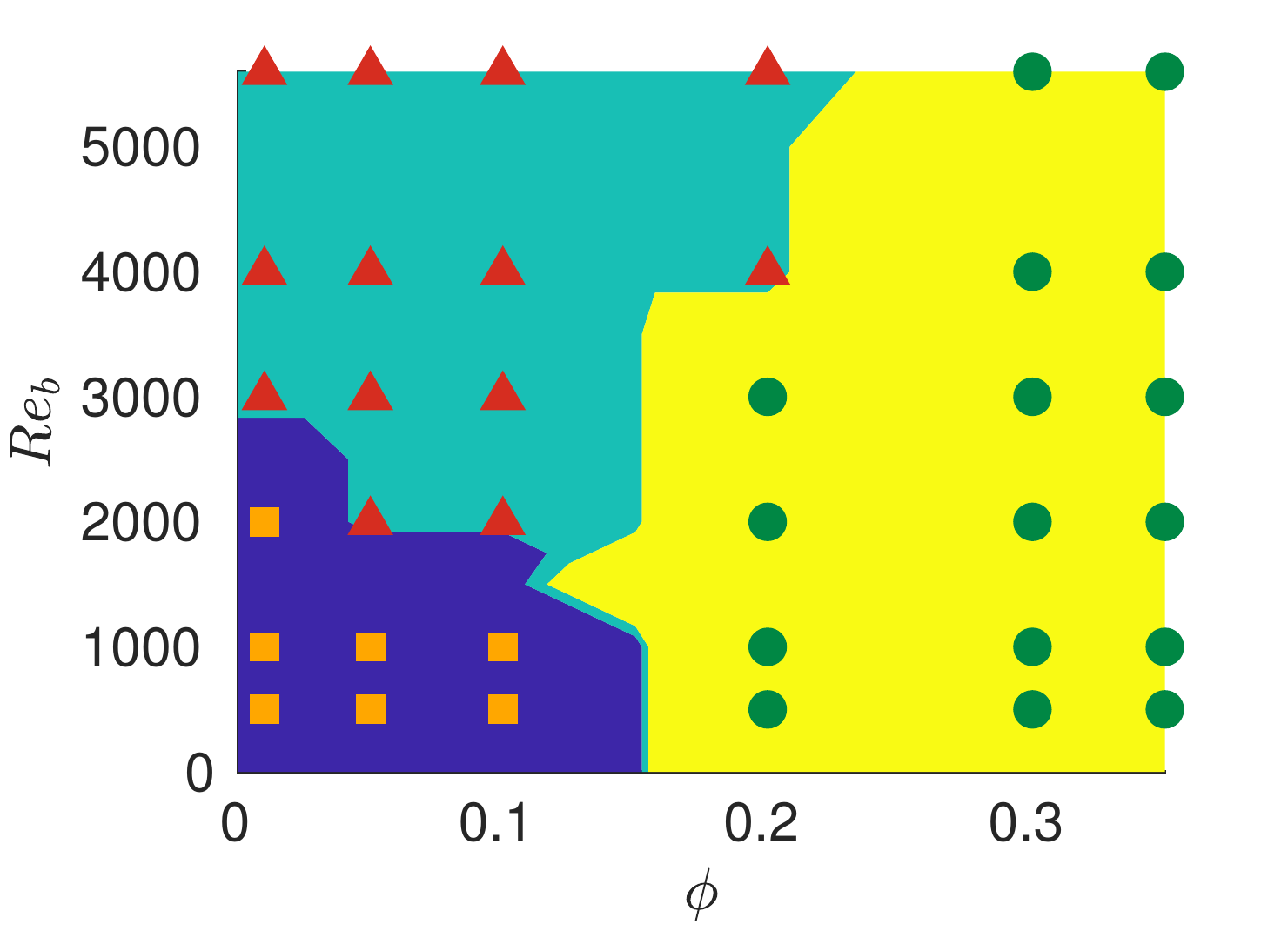}
   \put(-470,160){$(a)$}
   \put(-230,160){$(b)$}\\
   \caption{Maps of flow regimes in the Reynolds number--volume fraction plane,  identified by the dominant contribution to the momentum budget. Blue color: viscous dominated regime; green: turbulent-like flow; yellow: particle-stress dominated regime. The different symbols display the available simulation data. (a) Results from \cite{Lashgari-PRL-2014} and \cite{lashgari-IJMF-2016}. (b) Present simulations. }
   \label{fig:maps}
\end{figure}

In \cite{Lashgari-PRL-2014} and \cite{lashgari-IJMF-2016}, the total momentum transfer in suspensions is examined, as done here for the heat transfer, by a phase-ensemble average, see also \cite{picano-JFM-2015}.
The three contributions to the total stress $\tau (y)$ are the viscous stress $\tau_V$, the Reynolds stress $\tau_T$ (due to correlated motions of fluid and particles) and the particle stress $\tau_P$ (due to the particle stresslet and particle-particle interactions):
\begin{equation}
    \tau (y) = \tau_V + \tau_T + \tau_P  \mathrm{,}
  \label{eq:momentum_budget1}
\end{equation}
\begin{equation}
    \tau_V = \nu (1-\Phi) \frac{d U}{dy}  \mathrm{,}
  \label{eq:momentum_budget2}
\end{equation}
\begin{equation}
    \tau_T = -(1-\Phi) \langle u^{\prime} v^{\prime} \rangle - \Phi \langle u_p^{\prime} v_p^{\prime} \rangle  \mathrm{,}
  \label{eq:momentum_budget3}
\end{equation}
\begin{equation}
    \tau_P = \frac{\Phi}{\rho} \langle \sigma_{p \, xy} \rangle  \mathrm{,}
  \label{eq:momentum_budget4}
\end{equation}
where $\sigma_{p \, xy}$ is the general stress in the particle phase, projected in the streamwise direction. Based on this decomposition, three different regimes were identified, according to the transport mechanism mainly responsible for the momentum transfer: i) a viscous regime, at low Reynolds numbers and low particle concentrations; ii) a turbulent regime, dominated by the Reynolds stresses at high Reynolds and low concentrated and iii) a particulate regime, denoted as inertial shear-thickening, where the turbulent drag increases with the Reynolds number; however, the flow is dominated by the particulate stresses and turbulent fluctuations are secondary. The map with these 3 regimes is reported in figure~\ref{fig:maps}(a) where we replot the data in \cite{Lashgari-PRL-2014} and \cite{lashgari-IJMF-2016}.

The same momentum-budget analysis is performed for the simulations presented in this work, covering a similar range of Reynolds number and particle volume fraction, the only difference being the particle size: $2h/D=15$ in the present case and $2h/D=10$ in \cite{Lashgari-PRL-2014} and \cite{lashgari-IJMF-2016}, i.e.\ smaller particles in the present study.
The regime map pertaining this work is displayed in figure~\ref{fig:maps}(b), showing that for the smaller particle size, the onset of the particulate regime (inertial-sher thickening) shifts to higher volume fractions. Considering also the particle migration towards the channel centre, less pronounced in the case of smaller particles, we confirm that migration is proportional to the particle inertia and that, expectedly, the particle stress is also proportional to the particle size.

\section{Final remarks} \label{sec:Final_remarks}

\begin{figure}
  \centering
   \includegraphics[width=0.495\textwidth]{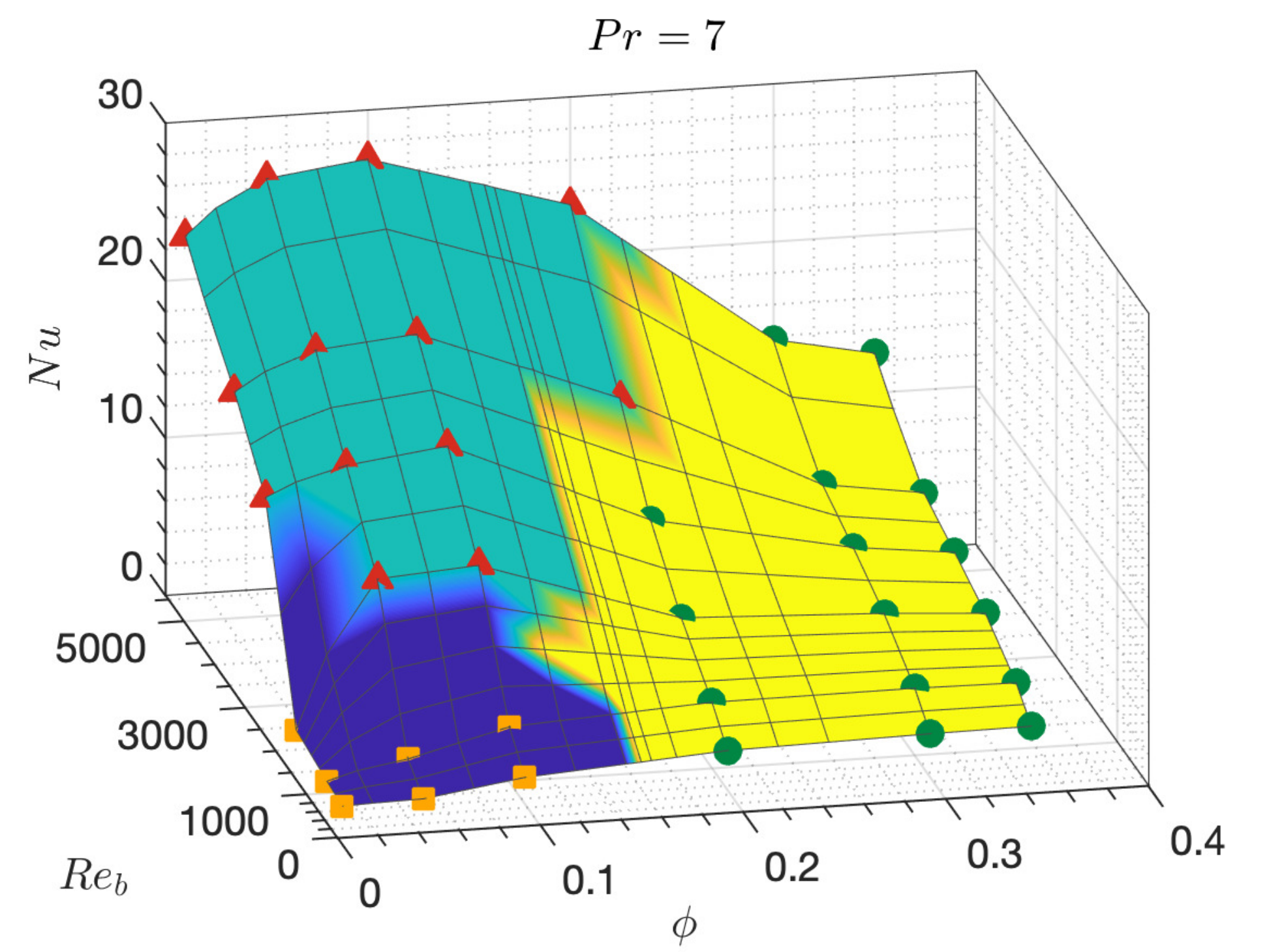}
   \includegraphics[width=0.495\textwidth]{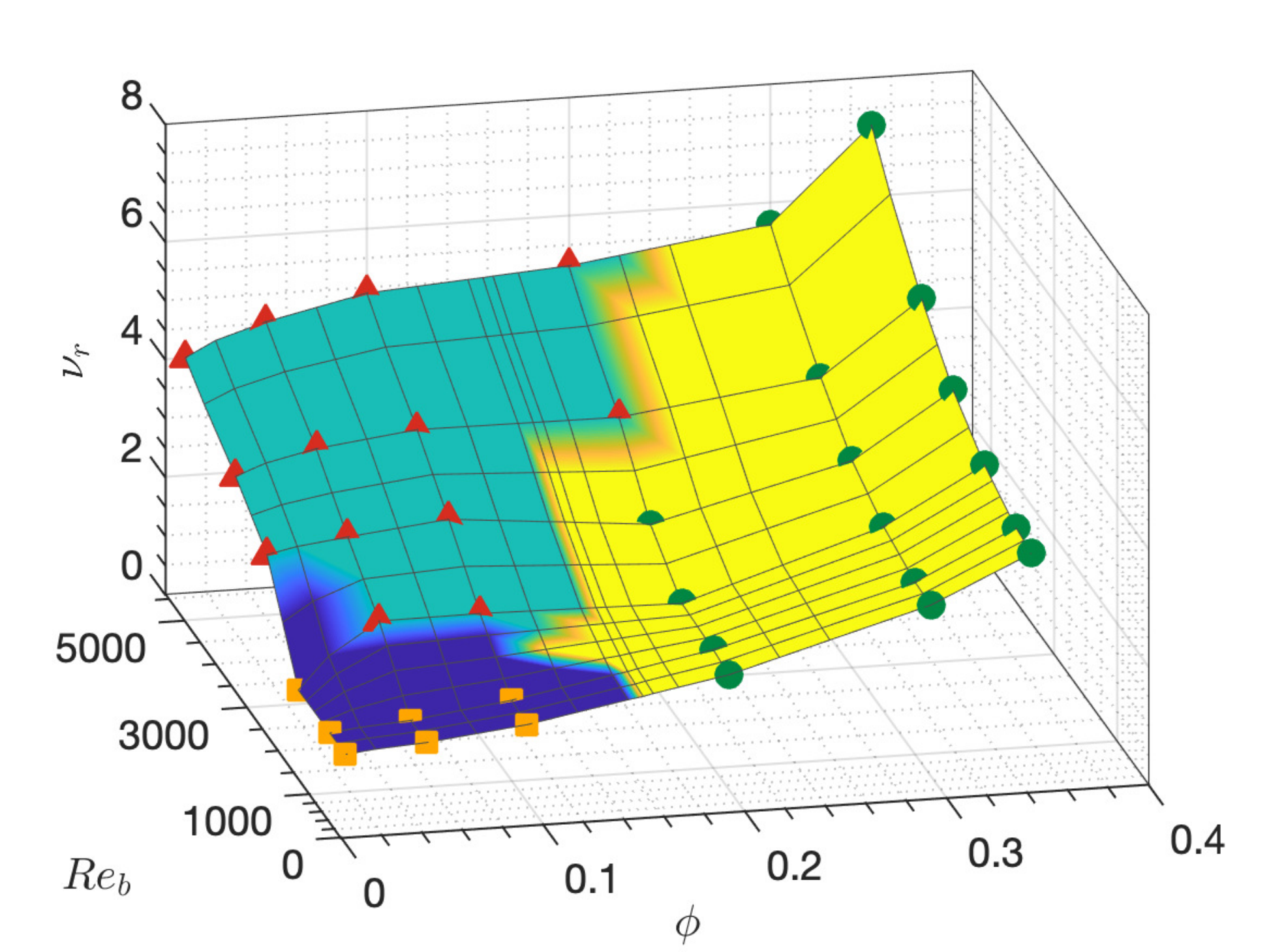}
   \put(-470,165){$(a)$}
   \put(-230,165){$(b)$}\\
   \includegraphics[width=0.495\textwidth]{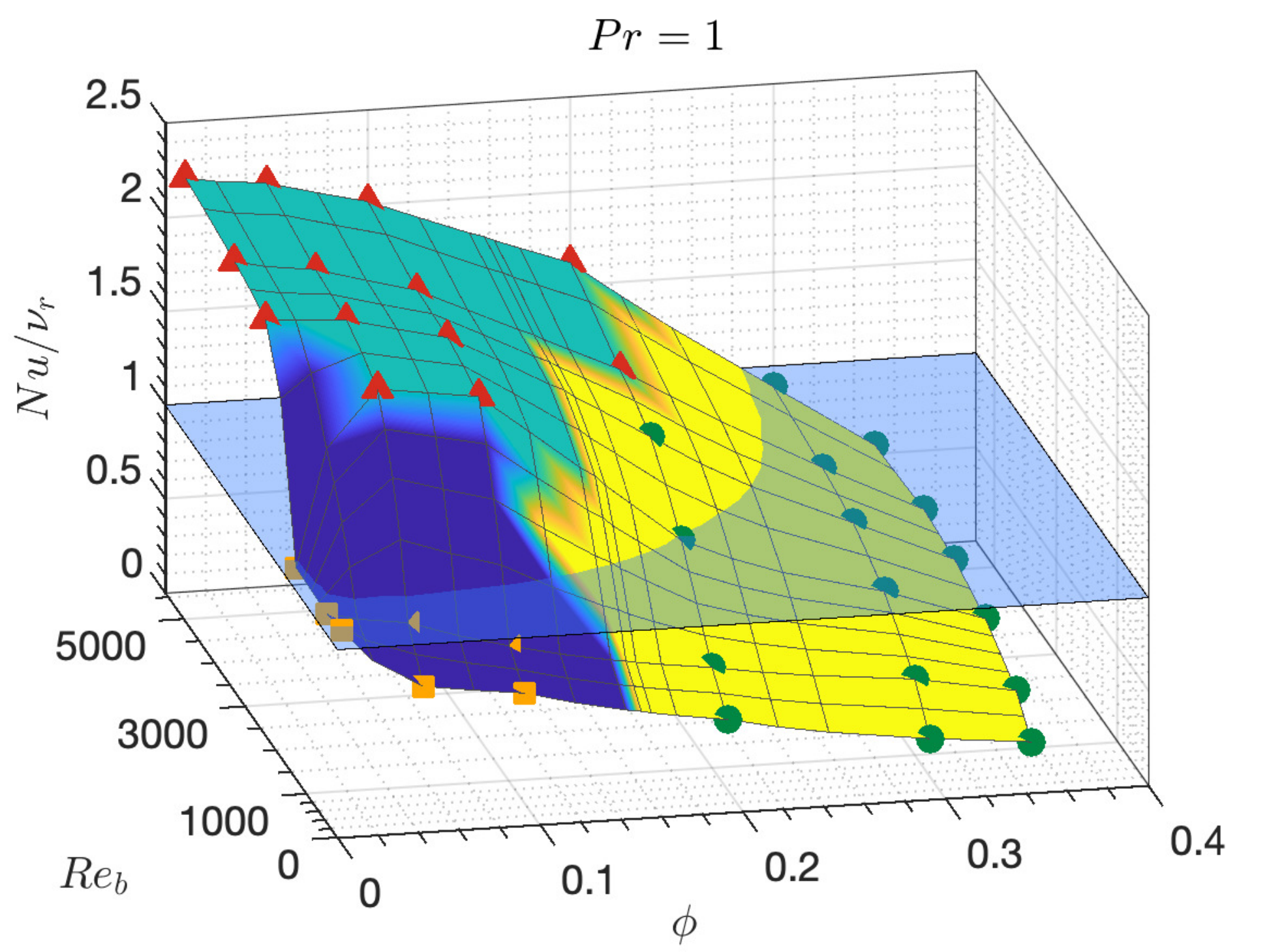}
   \includegraphics[width=0.495\textwidth]{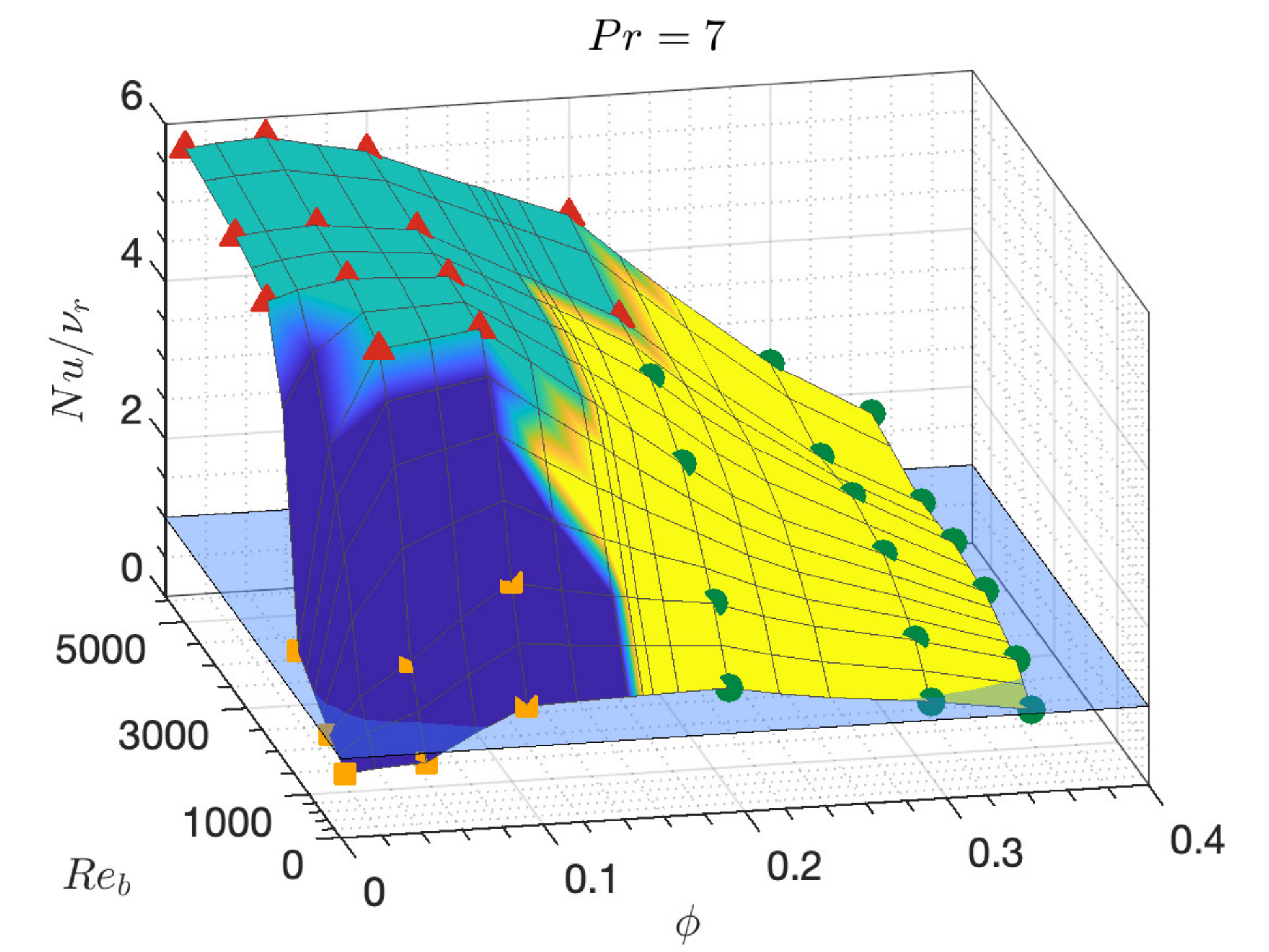}
   \put(-470,165){$(c)$}
   \put(-230,165){$(d)$}\\
   \includegraphics[width=0.495\textwidth]{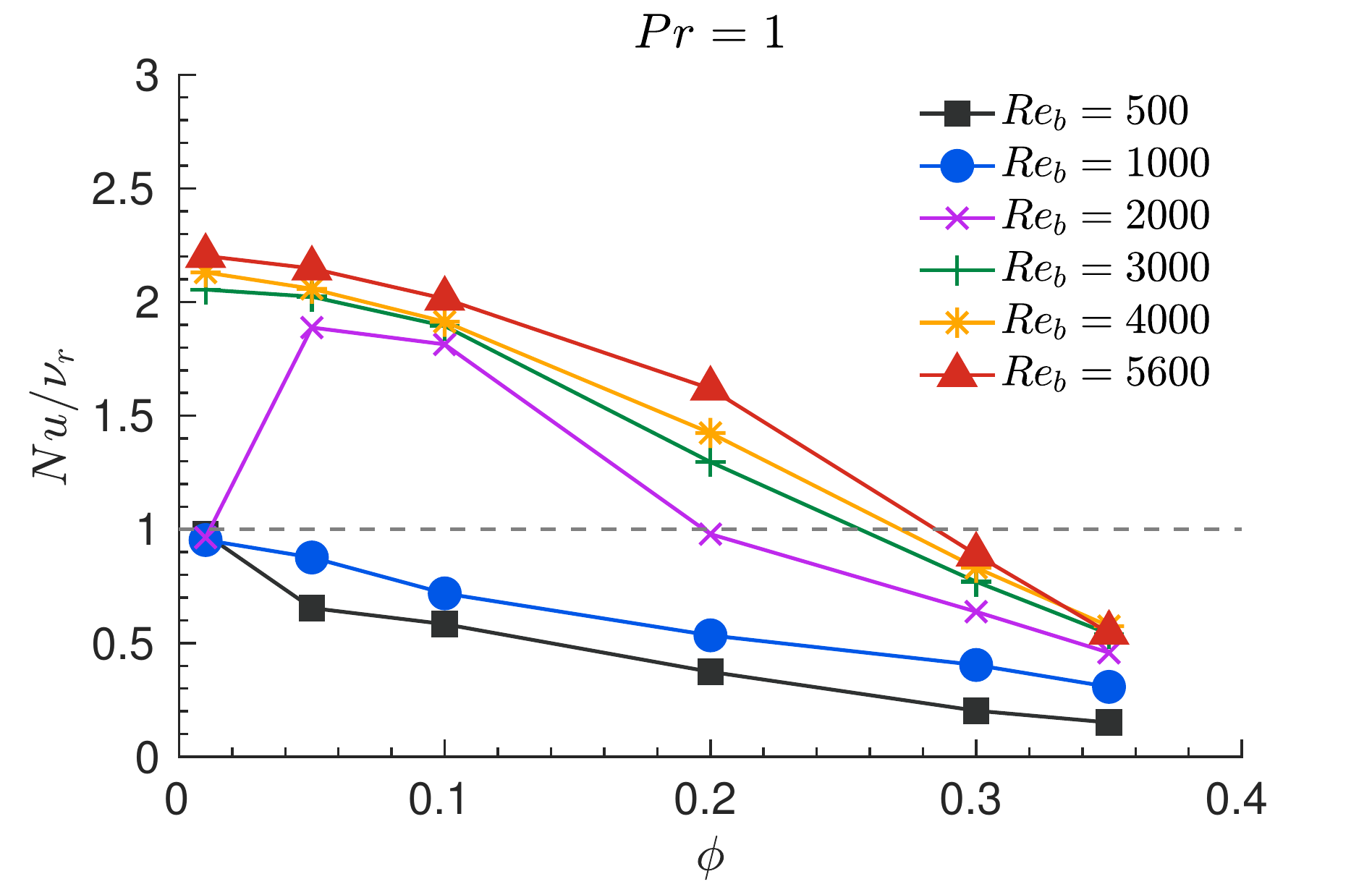}
   \includegraphics[width=0.495\textwidth]{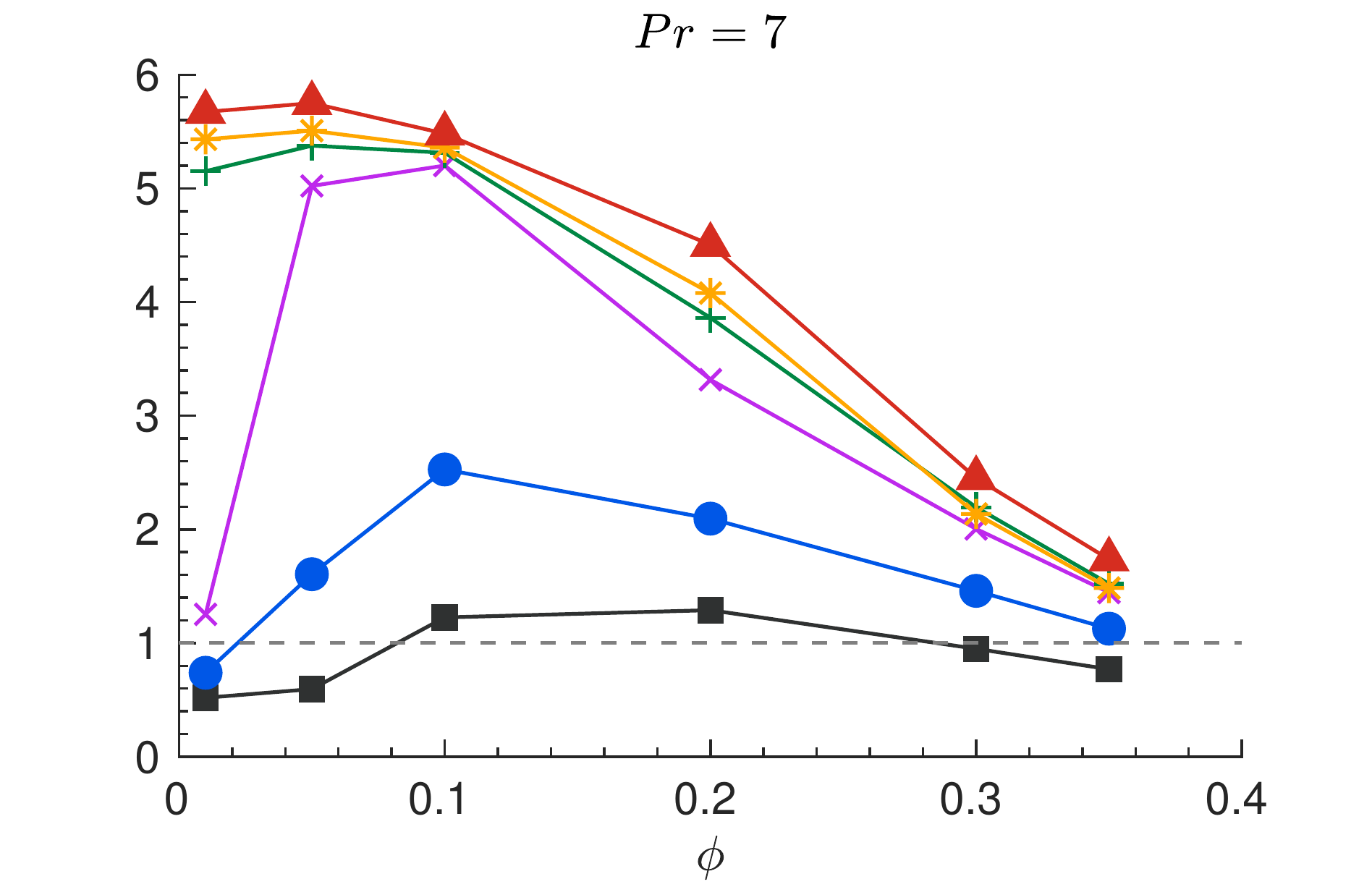}
   \put(-470,145){$(e)$}
   \put(-230,145){$(f)$}\\
   \caption{ Three-dimensional maps of the flow regimes in the Reynolds-volume fraction plane: (a) the Nusselt number $Nu$ for $Pr=7$, (b) the relative viscosity of the suspension $\nu_r$ and the ratio between the Nusselt number and the relative viscosity, $Nu / \nu_r$ for (c) $Pr=1$ and (d) $Pr=7$; the planes with light blue color mark the value $Nu / \nu_r = 1$. The color of the isosurface indicates the flow regime as shown in figure~\ref{fig:maps} and the symbols the simulation data available. (e): $Nu / \nu_r$ versus the volume fraction for the Prandtl number $Pr=1$ and (f) $Pr=7$; the dashed black line denotes the value $NU / \nu_r =1$.}
   \label{fig:3Dmap}
\end{figure}

We have presented results of simulations of heat transfer in suspensions of finite-size neutrally-buoyant spherical particles for solid volume fractions up to $35\%$ and bulk Reynolds numbers from 500 to 5600. To summarise the main findings, 
we present in figure~\ref{fig:3Dmap}(a) a three-dimensional map of the effective diffusivity as a function of the Reynolds number and the volume fraction, where the surface indicating heat transfer enhancement with respect to the laminar single-phase flows is colored by the corresponding flow regime, viscous, turbulent or inertial shear-thickening as identified in \cite{Lashgari-PRL-2014}.
We relate these 3 regimes to the heat transfer enhancement discussed in figure~\ref{fig:heat_enhancement} and in~\ref{fig:3Dmap}(a). 
In the viscous dominated regime, dark blue color as in figure~\ref{fig:maps}(b), 
heat transfer is mainly due to molecular diffusion, with a significant increase due to the particle induced  fluctuations in the flow. This can be related to inertial effects, as discussed in \cite{ardekani-JFM-2017}.
In the turbulent-like regime, upper-left corner in the regime map in green, we observe the largest enhancement of the global heat transfer, in this case dominated by the turbulent heat flux. Finally, in the particulate regime, right side of the map, the heat transfer enhancement decreases as mixing is quenched by the particle migration towards the channel core and by the formation of a compact loosely packed core region. In this regime, the contribution of molecular diffusion to the total heat transfer becomes again important, especially in this core region.

It should be noted that increasing the heat transfer efficiency by adding solid phase particles comes with the cost of increasing the effective viscosity of the suspension. This means a higher friction at the walls and higher external power needed to drive the flow.
To quantify this effect, we introduce the relative viscosity $\nu_r \equiv \nu_e / \nu$, with $\nu_e$  the effective viscosity of the suspension.
This is the viscosity of a single-phase laminar flow that would give the same shear stress as in our simulations:
\begin{equation}
    \nu_e = \tau_{xy} / (\rho \frac{dU}{dy} \bigg|_{lam. \, \phi=0\%})  \mathrm{.}
  \label{eq:eff_vis}
\end{equation}
The relative viscosity pertaining the cases under investigation, $\nu_r(Re,\phi)$, is depicted in a three-dimensional map in figure~\ref{fig:3Dmap}(b).
The data confirms that the effective viscosity of the suspension increases both with the volume fraction $\phi$ and the Reynolds number.
At low $Re_b$, it follows the classical empirical fits obtained in the limit of vanishing inertia, see e.g. \cite{stickel2005}.
In the laminar regime, the effective viscosity increases when  increasing the Reynolds number, which was explained by the microstructure anisotropy and by excluded volume effects in \cite{picano2013}. In turbulent flows, the drag increases as a consequence of the additional mixing introduced by the Reynolds stresses. On the other hand, the increased particle stresses, also associated to the migration towards the channel centreline discussed here,  are the responsible for the rapid increase of $\nu_r$ observed at the highest values of $\phi$, almost independently of the Reynolds number, see right-most points in figure~\ref{fig:3Dmap}(b).

Finally, figures~\ref{fig:3Dmap}(c-d) display the values of $Nu / \nu_r$ as a function of the bulk Reynolds number and of the volume fraction of the solid phase for Prandtl numbers $1$ and $7$ in three-dimensional plots, while the same data is depicted in two-dimensional plots in panels (e-f) for easier comparison.
The ratio between the Nusselt number and the effective viscosity of the suspension can be used to measure the efficiency of the heat transfer enhancement, as it accounts for the external power needed to drive the flow.
The light-blue color planes in the figure denotes the value $Nu / \nu_r  = 1$: for the points above this plane, the gain in heat transfer is more than the increase in the power needed to overcome the frictional forces, while the points below this plane indicate situations where the increase in heat transfer is lower than the increase in the effective viscosity. As depicted in figure~\ref{fig:3Dmap} (c) for $Pr=1$, only for volume fractions $\phi < 25\%$ and $Re_b > 2000$, so turbulent flows, the addition of particles leads to an increase of the heat transfer which exceeds the increase of the corresponding pressure power needed to drive the flow. On the other hand, for the Prandtl number $Pr=7$, the majority of the studied cases result in a value $Nu / \nu_r > 1$; this indicates that, for fluids with lower thermal diffusivity, the impact of increasing inertia or/and adding solid particles on the heat transfer enhancement compared to the drag increase is more significant than for fluids with higher values of thermal diffusivity.
At $Pr=7$, for each $Re_b$, we can identify an optimal volume fraction where the heat transfer enhancement is maximum, considering the drag increase constraint. This optimal value of the volume fraction decreases, increasing the Reynolds number, see figure~\ref{fig:3Dmap}(f).

To conclude, we have shown how interface-resolved simulations can provide physical insights into momentum and heat transfer in particle suspensions. In the future, one might extend this work by considering particles of different shape and a different heat conductivity in the solid and liquid phases.
The numerical methods proposed here can also be changed to consider mass transfer at the solid boundaries, to study e.g.\ gas absorption by a solid phase. In this case, it might be important to also model chemical reactions at the interface.

\section*{Declaration of Competing Interest}
The authors declared that there is no conflict of interest.

\section*{Acknowledgements}
This work was supported by the European Research Council grant no. ERC-2013-CoG-616186, TRITOS and by the Swedish Research Council grant no. VR 2014-5001.
The authors acknowledge computer time provided by SNIC (Swedish National Infrastructure for Computing) and by the National Infrastructure for High Performance Computing and Data Storage in Norway (project no. NN9561K).
We also thank TetraPak and Prof. Fredrik Innings for fruitful discussions within the STEM project.

\bibliography{bibfile}

\end{document}